\documentclass[10pt]{article}

\usepackage{amsmath}
\usepackage{amssymb}

\usepackage{graphicx}

\usepackage{cite}

\usepackage{color} 

\usepackage[labelfont=bf,labelsep=period,justification=raggedright]{caption}

\makeatletter
\renewcommand{\@biblabel}[1]{\quad#1.}
\makeatother

\date{}

\usepackage{amsthm}
\usepackage[utf8x]{inputenc}
\usepackage{setspace}
\usepackage{fullpage}
\usepackage{multirow}
\usepackage{subfigure}
\usepackage{amsmath}
\usepackage{enumerate}
\usepackage{lineno}
\usepackage{epsfig}

\newcommand{\eq}[1]{Eq.~(\ref{eq:#1})}

\newcommand{\eqs}[2]{Eqs.~(\ref{eq:#1}) and (\ref{eq:#2})}

\newcommand{\Eqss}[2]{Equations~(\ref{eq:#1}) through (\ref{eq:#2})}
\newcommand{\fig}[1]{Fig.~\ref{fig:#1}}

\newcommand*\patchAmsMathEnvironmentForLineno[1]{%
  \expandafter\let\csname old#1\expandafter\endcsname\csname #1\endcsname
  \expandafter\let\csname oldend#1\expandafter\endcsname\csname end#1\endcsname
  \renewenvironment{#1}%
     {\linenomath\csname old#1\endcsname}%
     {\csname oldend#1\endcsname\endlinenomath}}%
\newcommand*\patchBothAmsMathEnvironmentsForLineno[1]{%
  \patchAmsMathEnvironmentForLineno{#1}%
  \patchAmsMathEnvironmentForLineno{#1*}}%
\AtBeginDocument{%
\patchBothAmsMathEnvironmentsForLineno{equation}%
\patchBothAmsMathEnvironmentsForLineno{align}%
\patchBothAmsMathEnvironmentsForLineno{flalign}%
\patchBothAmsMathEnvironmentsForLineno{alignat}%
\patchBothAmsMathEnvironmentsForLineno{gather}%
\patchBothAmsMathEnvironmentsForLineno{multline}%
}

\begin{document}

\begin{flushleft}
{\Large
\textbf{Evolutionary Game Dynamics in Populations with Heterogenous Structures}
}
\\
Wes Maciejewski$^{1, \ast}$, 
Feng Fu$^{2}$, 
Christoph Hauert$^{1}$
\\
\bf{1} Department of Mathematics, The University of British Columbia, Vancouver, British Columbia, Canada
\\
\bf{2} Theoretical Biology, Institute of Integrative Biology, ETH Z\"{u}rich, Z\"{u}rich, Switzerland
\\
$\ast$ E-mail: wes@math.ubc.ca
\end{flushleft}

\linenumbers
\doublespacing

\section*{Abstract}

Evolutionary graph theory is a well established framework for modelling the evolution of social behaviours in structured populations.  An emerging consensus in this field is that graphs that exhibit heterogeneity in the number of connections between individuals are more conducive to the spread of cooperative behaviours. In this article we show that such a conclusion largely depends on the individual-level interactions that take place. In particular, averaging payoffs garnered through game interactions rather than accumulating the payoffs can altogether remove the cooperative advantage of heterogeneous graphs while such a difference does not affect the outcome on homogeneous structures. In addition, the rate at which game interactions occur can alter the evolutionary outcome. Less interactions allow heterogeneous graphs to support more cooperation than homogeneous graphs, while higher rates of interactions make homogeneous and heterogeneous graphs virtually indistinguishable in their ability to support 
cooperation.  Most importantly, we show that common measures of evolutionary advantage used in homogeneous populations, such as a comparison of the fixation probability of a rare mutant to that of the resident type, are no longer valid in heterogeneous populations. Heterogeneity causes a bias in where mutations occur in the population which affects the mutant's fixation probability. We derive the appropriate measures for heterogeneous populations that account for this bias.

\section*{Author Summary}

Understading the evolution of cooperation is a persistent challenge to evolutionary theorists. A contemporary take on this subject is to model populations with interactions structured as close as possible to actual social networks. These networks are heterogeneous in the number and type of contact each member has. Our paper demonstrates that the fate of cooperation in such heterogeneous populations critically depends on the rate at which interactions occur and how interactions translate into the fitnesses of the strategies. We also develop theory that allows for an evolutionary analysis in heterogeneous populations. This includes deriving appropriate criteria for evolutionary advantage. 

\section*{Introduction}

Population structure has long been known to affect the outcome of an evolutionary process \cite{wright:Genetics:1931, kimura:Genetics:1964b, levins:BSEA:1969, nowak:nature:1992b}. Evolutionary graph theory has emerged as a convenient framework for modelling structured populations \cite{nowak:nature:1992b, lieberman:nature:2005}. Individuals reside on vertices of the graph and the edges define the interaction neighbourhoods. 

A variety of processes have been investigated on a number of graph classes. However, few analytical results exist in general, since an arbitrary graph may not exhibit sufficient symmetry to aid calculations. The most general class of graphs for which analytical results are known is the class of homogeneous (\emph{vertex-transitive}) graphs. Such a graph $G$ has the property that for any two vertices $v_i$ and $v_j$ there exists a structure-preserving transformation $g$ of $G$ such that $g(v_i)=v_j$. It is worth noting that not all regular graphs are homogeneous; an extreme example is the Frucht graph \cite{frucht:CJM:39}, which is regular of degree $3$ and has only the trivial symmetry. Intuitively, this class consists of graphs that ``look'' the same from any vertex. The amount of symmetry in such graphs has allowed for a complete set of analytical results for restricted types of evolutionary processes and weak 
selection \cite{ohtsuki:PRSB:2006, taylor:Nature:2007, grafen:JTB:2008}. Despite the tractability of calculations on homogeneous graphs, natural population structures are seldom homogeneous. Therefore it is important to understand the effects of \emph{heterogeneous} population structures on evolutionary processes \cite{nowak:nature:1992b, taylor:Nature:2007, ohtsuki:Nature:2006} and, in particular, on the evolution of cooperation. 

In the simplest case there are two strategic types: cooperators that provide a benefit $b$ to their interaction partner at some cost $c$ to themselves ($b>c>0$), whereas defectors provide neither benefits nor incur costs. This basic setup is known as an instance of the prisoner's dilemma and reflects a conflict of interest because mutual cooperation yields payoff $b-c>0$ and hence both parties prefer this outcome over mutual defection, which yields a payoff of zero. However, at the same time each party is tempted to defect in order to avoid the costs of cooperation. The temptation of increased benefits for unilateral defection thwarts cooperation -- to the detriment of all. This conflict of interest characterizes social dilemmas \cite{dawes:ARP:1980,hauert:JTB:2006a}.

\begin{center}
\begin{table}[h!]
\centering

\begin{tabular}{c | c | c|}
& $A$ & $ B$ \\
\hline 
$A$ & $1$ & $S$ \\
\hline 
$B$ & $T$ & $0$ \\
\hline
\end{tabular}\\

\medskip

\caption{The payoff matrix for a general $2$ by $2$ strategy game. Here $S$ and $T$ are real numbers.}
\end{table}
\end{center}

More general kinds of interactions between two individuals and two strategic types, $A$ and $B$, can be represented in the form of a $2\times 2$ payoff matrix as in Table 1. The payoffs garnered from these game interactions affect an individual's expected number of offspring by altering their propensity to have offspring (their \emph{fitness}) or their \emph{survival}. The expected number of offspring is determined by the fitness of the individuals and some population updating process, which will be made precise in the next section. The offspring produced during the population update have the potential to change the strategy composition of the population. An increase in the abundance of one strategy over a sufficiently large time scale indicates that strategy is favoured by evolution.

It can be shown, for replicator dynamics, for example \cite{taylor:MB:1978, hofbauer:book:1998}, that any payoff matrix can be reduced to the matrix in Table 1 without loss of generality because adding a constant term to the payoff matrix does not affect the dynamics and multiplying the payoffs by a positive factor merely rescales the time. Therefore we can always shift the payoffs such that $B$-$B$-encounters return a payoff of zero and scale all other payoffs such that $A$-$A$-encounters yield a payoff of $1$. In the Accumulated versus Averaged Payoffs section we show that the generality of the matrix in Table 1 extends to other forms of stochastic dynamics in finite populations based on the frequency dependent Moran process \cite{nowak:Nature:2004}.

The (additive) prisoner's dilemma introduced before corresponds to the special case with $S=-c/(b-c)$ and $T=b/(b-c)$. Rescaling the payoff matrix in Table 1 by $b-c$ yields the traditional form, Table 2. More generally, the prisoner's dilemma requires $S<0$ and $T>1$ to result in the characteristic conflict of interest outline above. The special case of the additive prisoner's dilemma, Table 2, effectively reduces the game to a single parameter with $T=1-S$ (and $S<0$). Moreover it has the special property that when an individual changes its strategy, the payoff gain (or loss) is the same, regardless of the opponents' strategy -- the so-called \emph{equal-gains-from-switching} property \cite{nowak:AAM:1990}.

\begin{center}
\begin{table}[h!]
\centering

\begin{tabular}{c | c | c|}
& $C$ &  $D$ \\
\hline 
$C$ & $b-c$ & $-c$ \\
\hline 
$D$ & $b$ & $0$ \\
\hline
\end{tabular}\\

\medskip

\caption{The payoff matrix for an additive prisoner's dilemma game.}
\end{table}
\end{center}

In the absence of structure, cooperators dwindle and disappear in the prisoner's dilemma. In contrast, structured populations enable cooperators to form clusters, which ensures that cooperators more frequently interact with other cooperators than they would with random interactions \cite{van-baalen:JTB:1998, hauert:PRSB:2001}. Such assortment between cooperators is essential for the survival of cooperation \cite{fletcher:PRSB:2009}. 

In heterogeneous graphs not all vertices have the same number of connections and hence the fitnesses of individuals may be based on different numbers of interactions. Because of this, some vertices are more advantageous to occupy than others. However, which sites are favourable depends on the type of population dynamics. In particular, for the Moran process in structured populations it is important to distinguish between birth-death and death-birth updating \cite{ohtsuki:Nature:2006,zukewich:PlosOne:2013,maciejewski:JTB:2013}, i.e. whether first an individual is randomly selected for reproduction with a probability proportional to its fitness and then the clonal offspring replaces a (uniformly) randomly selected neighbour -- or, if first an individual is selected at random to die and then the vacant site is repopulated with the offspring of a neighbouring individual with a probability proportional to its fitness. Even in homogenous populations the sequence of events is of crucial importance but becomes 
even 
more pronounced in heterogenous structures \cite{ohtsuki:Nature:2006,zukewich:PlosOne:2013}. 

In order to illustrate that the population dynamics may bestow an advantage to individuals occupying certain sites in a heterogeneous population, consider neutral evolution, where game payoffs do not affect the evolutionary process and all individuals have the same fitness. For birth-death updating every individual is chosen to reproduce with the same probability but neighbours of individuals with few connections are replaced more frequently. Hence vertices with fewer neighbours are more favourable than those with many connections. Conversely, for death-birth updating every individual has the same expected life time but highly-connected individuals, or, \emph{hubs}, get more frequently a chance to produce offspring, since one of their many neighbours dies, and are thus more favourable than vertices with few neighbours \cite{broom:JSTP:2011,li:PLOS:2013, maciejewski:JTB:2013}. A simple example of this is a $3$-line graph, one central vertex connected to two end vertices. In the birth-death process, the 
central vertex is replaced with probability $2/3$, while either end vertex is replaced with probability $1/6$, while in the death-birth process, the central vertex \emph{replaces} either end vertex with probability $2/3$ and either end replaces the centre with probability $1/6$ \cite{maciejewski:JTB:2013}. The upshot is, even though the fitness of all individuals is the same, the effective number of offspring produced depends on the 
dynamics as well as an individual's location in the population.

The intrinsic advantage of some vertices over others can be further enhanced through game interactions leading to differences in fitness that depend on an individual's strategy as well as its position on the graph. For example, a cooperator occupying a favourable vertex can more easily establish a cluster of cooperators, which creates a positive feedback through mutual increases in fitness. Conversely, a favourable vertex also supports the formation of a cluster of defectors but this results in a negative feedback and lowers the fitness of the defector in the favourable vertex. The fact that heterogeneity can promote cooperation was first observed for the prisoner's dilemma and snowdrift games \cite{santos:PRL:2005, santos:PRSB:2006} and has subsequently been confirmed for public goods games \cite{santos:PlosCB:2006, santos:Nature:2008}. However, the detailed effects not only crucially depend on the dynamics but also on how fitnesses are determined. For example, heterogenous population structures 
favour cooperation if payoffs from game interactions are accumulated but that advantage disappears if payoffs are averaged \cite{tomassini:IJMP:2007, szolnoki:PA:2008, antonioni:ACS:2012}. 

The effects of population structure on the outcome of evolutionary games is sensitive to a number of factors: population dynamics \cite{huberman:PNAS:1993,ohtsuki:Nature:2006,zukewich:PlosOne:2013}, translation of payoffs into fitness \cite{masuda:PRSB:2007a,tomassini:IJMP:2007, perc:PRE:2008a, pacheco:PLoSCB:2009,grilo:JTB:2011,antonioni:ACS:2012} and the type of game played -- for example, spatial structure tends to support cooperation in the prisoner's dilemma but conversely, in the snowdrift game, spatial structure may be detrimental \cite{hauert:Nature:2004}. Macroscopic features of the evolutionary process on the level of the population, such as frequency and distribution of cooperators, are determined by microscopic processes on the level of individuals. In the current article, we discuss some of these microscopic processes, such as averaging and accumulating payoffs, and the rate at which interactions take place, and illustrate how they affect an evolutionary outcome. We also develop a general 
framework to determine 
evolutionary advantage in finite, heterogeneous populations.

The manuscript is organized as follows. Sections ``Accumulated and Averaged Payoffs'' and ``Criteria for Evolutionary Success'' largely review the literature concerning evolution on heterogeneous graphs, though we extend existing results to general 2 by 2 games and focus on an immitation process. Interspersed in these sections are new observations and results (eg. the criteria for evolutionary success section) that aid in establishing a consistent framework on which we base our main results presented in the section ``Stochastic Interactions and Updates''.

\section*{Results}

\section*{\label{sect:accuavg}Accumulated versus Averaged Payoffs}
In heterogenous population structures individuals naturally engage in different numbers of interactions. This renders comparisons of the performances of individuals more challenging. One natural approach is to simply accumulate the game payoffs. This clearly puts hubs with many neighbours in a strong position as scoring many times even a small payoff may still exceed few large payoffs. To avoid this bias in favour of hubs, game payoffs can be averaged. Interestingly, these two approaches not only play a decisive role for the evolutionary outcome but also entail important biological implications.

Consider two different ways to translate the total, accumulated payoffs $\pi_i$ of an individual $i$ into its fitness $f_i$:
\begin{subequations}
\label{eq:fecund}
\begin{align}
\label{eq:accufecund}
f_i &= e^{\delta\pi_i},&&\text{\footnotesize\emph{accumulated}}&\\
\label{eq:avefecund}
f_i &= e^{\delta\frac{\pi_i}{n_i}},&&\text{\footnotesize\emph{averaged}}&
\end{align}
\end{subequations}
where $\delta>0$ denotes the strength of selection and $n_i$ is the number of interactions experienced by $i$. The limit $\delta\to0$ recovers the neutral process, where selection does not act. Note that the payoff matrix in Table 1 can still be used without loss of generality because adding a constant $\kappa$ merely changes the (arbitrary) baseline fitness from $1$ to $e^{\delta\kappa}$ and multiplying the payoffs by $\lambda$ is identical to simply changing the selection strength to $\delta\lambda$.

The exponential form of fitness in the above equations is mathematically convenient since it guarantees that the fitness is always positive, irrespective of the strength of selection and payoff values. It is worth noting that if the strength of selection is weak, that is, $\delta\ll1$, then 
\begin{subequations}
\begin{align}
f_i &= e^{\delta\pi_i} \approx 1 + \delta\pi_i + O(\delta),&&\text{\footnotesize\emph{accumulated}}&\\
f_i &= e^{\delta\frac{\pi_i}{n_i}} \approx 1 + \delta\frac{\pi_i}{n_i} + O(\delta),&&\text{\footnotesize\emph{averaged}}&
\end{align}
\end{subequations}
which represents another common form for fitness found in the literature \cite{taylor:Nature:2007}.

\subsection*{Homogenous populations}
In the past, details of the payoff accounting have received limited attention, or the two approaches have been used interchangeably,  because they yield essentially the same results for traditional models of spatial games, which focus on lattice populations \cite{nowak:nature:1992b, hauert:IJBC:2002} or, more generally, on homogenous populations \cite{szabo:PR:2007,taylor:Nature:2007,ohtsuki:Nature:2006}. In fact, the difference in payoff accounting reduces to a change in the selection strength because in homogenous populations each individual has the same degree $d_i=d$ (number of neighbours) and hence, on average, the same number of interactions $\bar n$ per unit time. If each individual interacts with all its neighbours then $\bar n=d$. Thus, the only difference is that the selection strength for accumulated payoffs is $\bar n$-times as strong as for averaged payoffs.

Therefore, in homogenous populations all individuals engage in the same number of interactions per unit time and consequently accumulating or averaging payoffs merely affects the strength of selection. Naturally, the converse question arises -- are uniform interaction rates restricted to homogenous graphs? Or, more generally, which class of graphs supports uniform interaction rates?

To answer this question, let us consider an arbitrary graph $G$ with adjacency matrix $W=[w_{ij}]$ where $w_{ij}\geq0$ indicates the weight or the strength of the (directed) edge from vertex $i$ to $j$. $w_{ij}>0$ if vertex $i$ is connected to $j$ and $w_{ij}=0$ if it is not. For example, the natural choice for the edge weights on undirected graphs is $w_{ij}=1/d_i$. That is, all $d_i$ edges leaving vertex $i$ have the same weight and hence $\sum_{j=1}^N w_{ij}=1$ for all $i$.

An individual on vertex $i$ is selected to interact with vertex $j$ with a probability proportional to $w_{ij}$. In this case we say vertex $i$ has initiated the interaction. Interactions with self are excluded by requiring $w_{ii}=0$. If there are $M$ interactions per unit time, then the average number of interactions $n_i$ that vertex $i$ engages in is given by
\begin{align}
n_i &= M\ \dfrac{\sum_{j=1}^N w_{ij}+\sum_{j=1}^N w_{ji}}{\sum_{j,k=1}^N w_{jk}},
\end{align}
where the fraction indicates the probability that vertex $i$ participates in one particular interaction either by initiating it (first sum in numerator) or initiated by neighbours of $i$ (second sum in numerator). On average each individual engages in $2M/N$ interactions. Note that the factor $2$ enters because each interaction affects two individuals. Therefore, a graph structure results in uniform interaction rates if and only if
\begin{align}
\label{eq:eqinter}
\dfrac{\sum_{j=1}^N \left(w_{ij}+ w_{ji}\right)}{\sum_{j,k=1}^N w_{jk}}&=\frac2N
\end{align}
holds for every vertex $i$, or equivalently, if $\sum_{j=1}^N (w_{ij}+ w_{ji})=C_0$ for all $i$ where $C_0$ is an arbitrary positive constant.

\begin{figure}[tbhp]
\centering
\includegraphics[width=0.6 \textwidth]{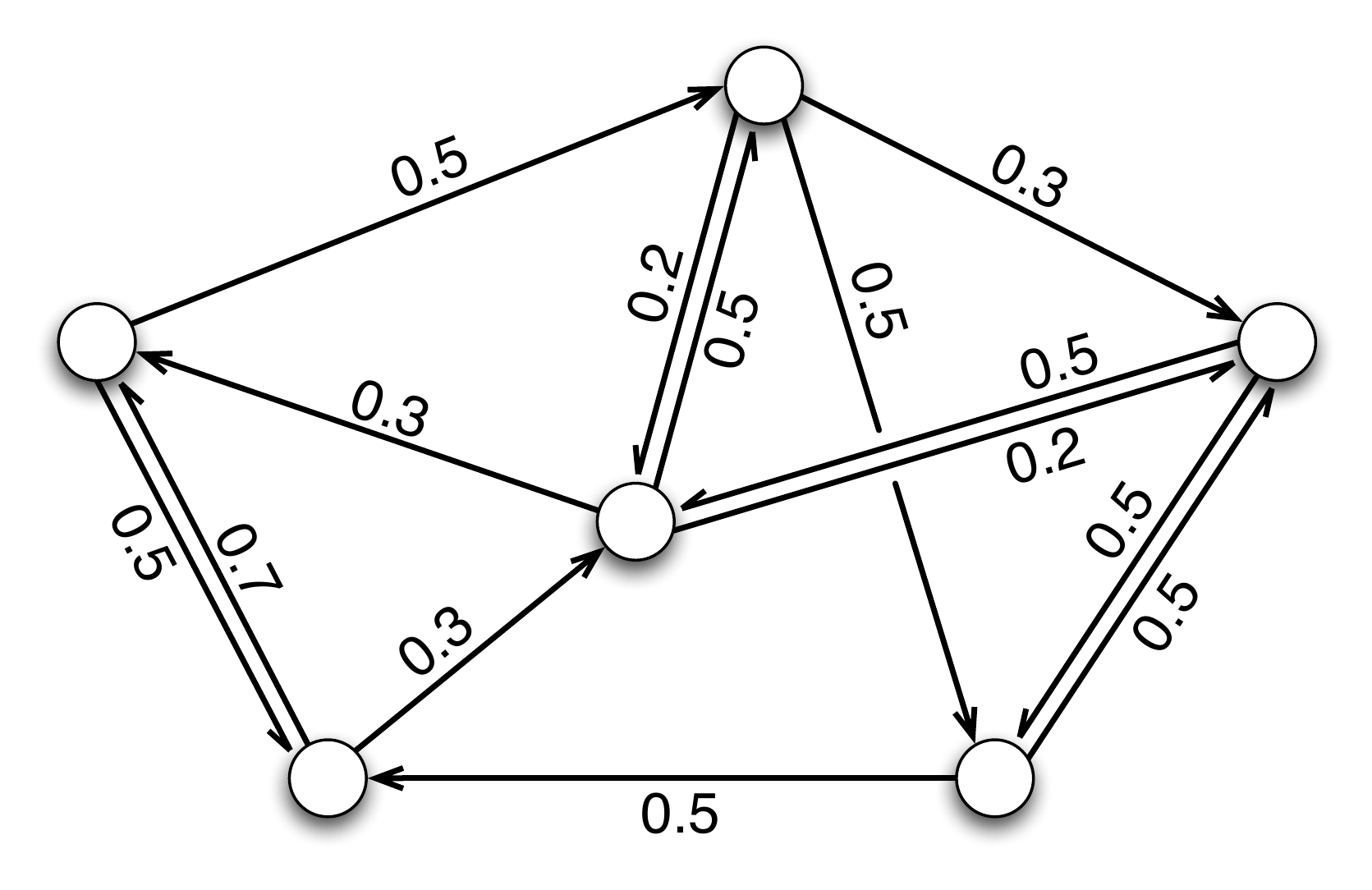}
\caption{\label{fig:isograph}A representative example of the broad class of circulation graphs. Note that the weights of edges entering as well as those leaving any vertex all sum to $1$.
}
\end{figure}

If the sum of the weights of all $d_i$ edges leaving vertex $i$, $\sum_{j=1}^N w_{ij}=C_1>0$, is the same for all $i$ then $\sum_{j,k=1}^N w_{kj}=N\cdot C_1$ and \eq{eqinter} requires that the sum of the weights of all incoming edges, $\sum_{j=1}^N w_{ji}=C_1$, for all $i$, as well to ensure uniform interaction rates. The class of graphs that satisfies the condition $\sum_{j=1}^N w_{ij}=\sum_{j=1}^N w_{ji}=C_1$ for all $i$ are called \emph{circulations} \cite{lieberman:nature:2005} and, in the special case with $C_1=1$, the adjacency matrix $W$ is doubly stochastic such that each row and column sums to $1$. A more generic representative of the broad class of circulation graphs is shown in \fig{isograph} but this does not include heterogenous graphs such as scale-free networks.
In order to illustrate that the number of interactions experienced by an individual depends on which vertex they reside, consider an arbitrary, random, undirected graph and assume that the degrees of adjacent vertices are uncorrelated. Under this assumption the approximate probability that vertices $i$ and $j$ are connected by an edge is
\begin{align}
w_{ij} = \frac{d_i\,d_j}{\bar d\,N},
\end{align}
where $\bar d=\sum_{i=1}^N d_i/N$ is the average vertex degree. Inserting into \eq{eqinter} yields
\begin{align}
n_i &= M\frac2N\frac{d_i}{\bar d}.
\end{align}
Hence, the number of interactions of one vertex scales linearly with its degree.

\begin{figure}[tbhp]
\centering
\includegraphics[ width=0.6 \textwidth]{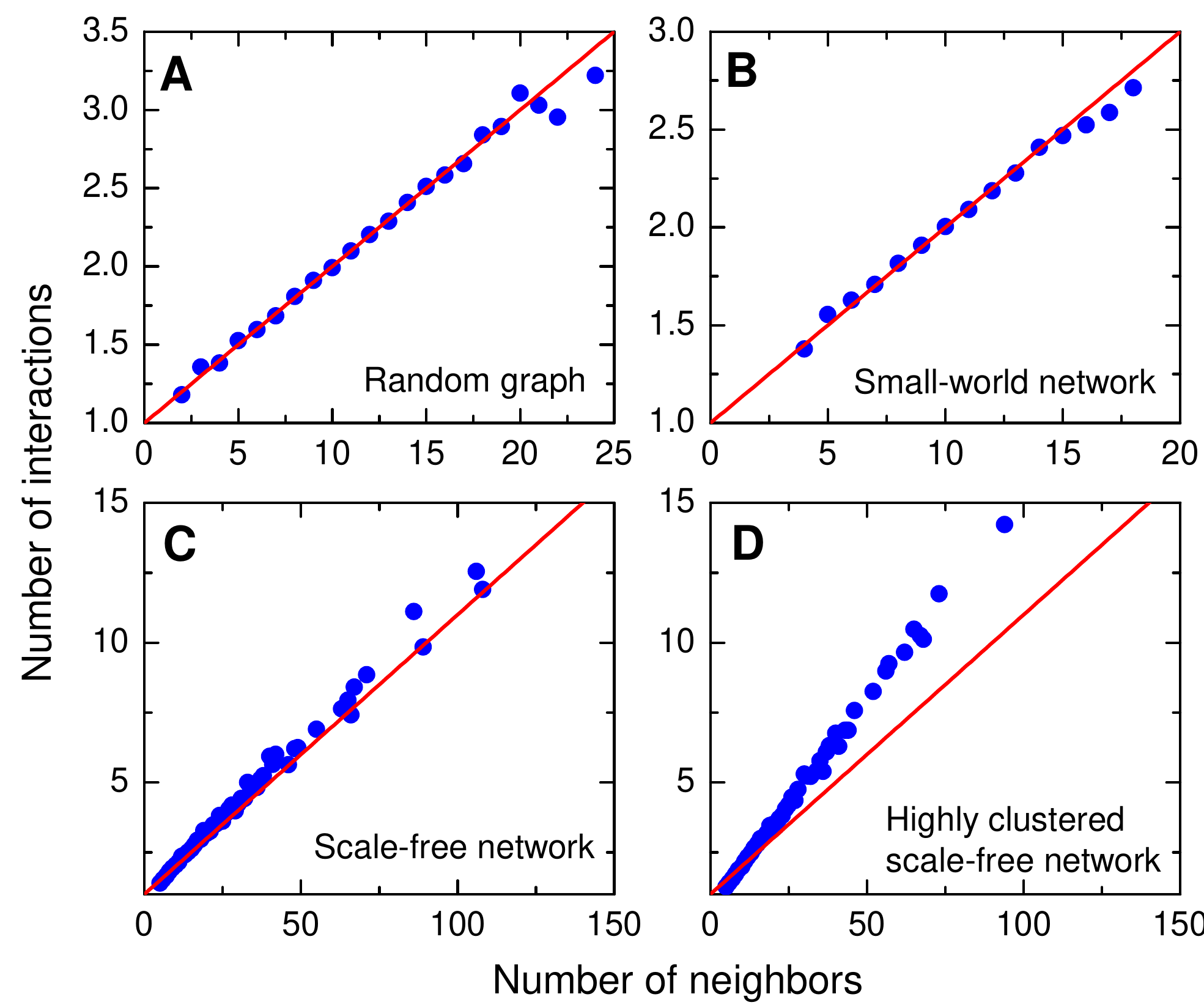}
\caption{\label{fig:numofint}Average number of interactions as a function of the degree of the vertex for different types of random heterogenous population structures: 
(A) Erd\'{o}s-R\'{e}nyi random graphs \cite{erdos:PMHAS:1960}, 
(B) Newman-Watts small-world networks \cite{newman:PRE:1999}.
(C) Barab\'{a}si-Albert scale-free networks \cite{barabasi:Science:1999}, and 
(D) Klemm-Eguiluz highly-clustered scale-free networks \cite{klemm:PRE:2002a}. 
All graphs have size $N=1000$ and an average degree of $\bar d = 10$. At each time step a randomly chosen individual interacts with a randomly selected neighbour. The average number of interactions is shown for simulations (blue dots) and an analytical approximation for graphs where the degrees of adjacent vertices are uncorrelated (red line, see \eq{nidi}).}
\end{figure}

Similarly, each vertex can initiate the same number of interactions, $m$. Then, with probability $w_{ji}/d_j$ the neighbouring vertex $j$ initiates an interaction with $i$:
\begin{align}
\label{eq:nidi}
n_i &= m\left(1+\sum_{j=1}^N\frac{w_{ji}}{d_j}\right) = m\left(1+\sum_{j=1\atop j\neq i}^N\frac1{d_j}\frac{d_i\,d_j}{\bar d\,N}\right) = m\left(1+\frac{N-1}N\frac{d_i}{\bar d}\right).
\end{align}
Again, vertices with a degree greater (less) than the average degree are expected to have more (fewer) interactions than on average. Interaction rates on various heterogenous networks are shown in \fig{numofint}.
This indicates that on undirected graphs uniform interaction rates can be achieved only on regular graphs, where all vertices have the same number of neighbours.

\subsection*{Heterogenous populations}
In recent years the focus has shifted from homogenous populations to heterogenous structures and, in particular, to small-world or scale-free networks because they capture intriguing features of social networks \cite{barabasi:Science:1999}. On these structures the accounting of payoffs becomes important and, indeed, a crucial determinant of the evolutionary outcome. If payoffs are accumulated, heterogenous structures further promote the evolution of cooperation \cite{santos:PRL:2005,santos:PRSB:2006,santos:Nature:2008, santos:JTB:2012}. In contrast, averaging the game payoffs can remove the ability for scale-free graphs to sustain higher levels of cooperation \cite{tomassini:IJMP:2007, szolnoki:PA:2008, antonioni:ACS:2012}. 

So far our discussion has focussed on interactions between individuals and the translation of payoffs into fitness. The next step is to specify how differences in fitness affect the population dynamics. The most common updating rules in evolutionary games on graphs fall into three categories: Moran birth-death and death-birth, and imitation processes. The evolutionary outcome can be highly sensitive to the choice of update rule. For example, supposing weak selection, cooperation in the prisoner's dilemma may only thrive under death-birth but not under birth-death updating \cite{ohtsuki:Nature:2006, taylor:Nature:2007, zukewich:PlosOne:2013}. 

In heterogenous populations the range of payoffs depends on the payoff accounting: if payoffs are averaged, the range is determined by the maximum and minimum values in the payoff matrix but if payoffs are accumulated the range additionally depends on the size and structure of the population. In particular, this difference may also affect the updating rule: for example, the pairwise comparison process $1/2+(f_j-f_i)/\alpha$ represents the probability that vertex $i$ adopts the strategy of vertex $j$ based on their fitnesses of $f_i$, $f_j$, respectively \cite{traulsen:PRL:2005,traulsen:PRE:2012}. This represents an imitation process where $\alpha$ denotes a sufficiently large normalization constant to ensure that the expression indeed remains a probability. Since $\alpha$ needs to be at least twice the range of possible fitness values, a generic choice of $\alpha$ becomes impossible for accumulated payoffs. 

\begin{figure}[thbp]
\centerline{\epsfig{file=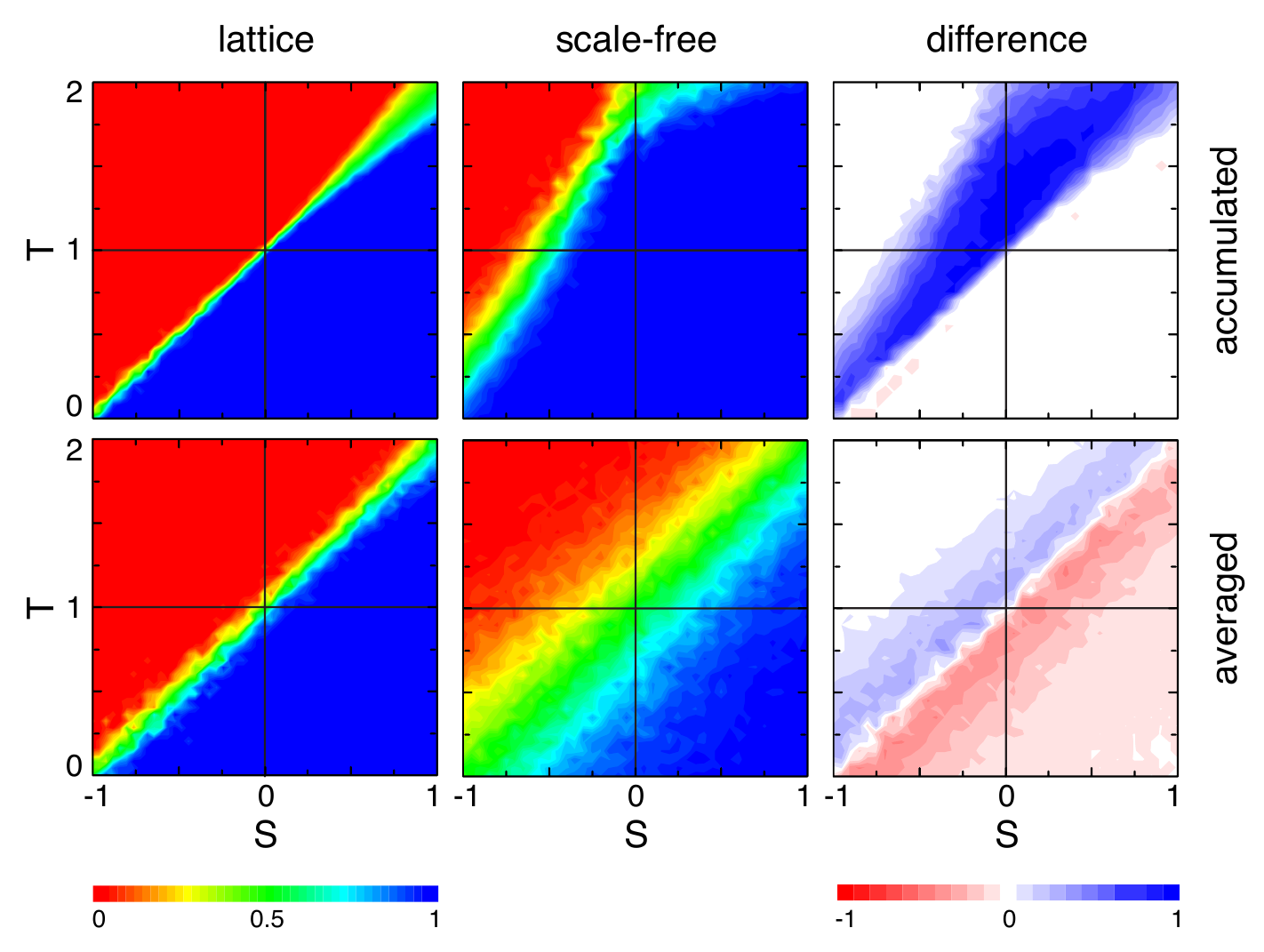,width=\textwidth}}
\caption{\label{fig:avgaccu}Average fraction of strategy $A$ for accumulated (top row) versus averaged (bottom row) payoffs in homogenous (left column) and heterogeneous (middle column) populations as well as the difference between them (right column) as a function of the game parameters $S$ and $T$ (see Table 1). In each panel the four quadrants indicate the four basic types of generalized social dilemmas: prisoner's dilemma (upper left), snowdrift or co-existence games (upper right), stag hunt or coordination games (lower left) and harmony games (lower right). Homogenous populations are represented by $50\times50$ lattices with von Neumann neighbourhood (degree $d=4$) and heterogenous populations are represented by Barab\'asi-Albert scale-free networks (size $N=2500$, average degree $\bar d=4$). The population is updated according to the imitation rule \eq{imitation}. The colours indicate the equilibrium fraction of strategy $A$ (left and middle columns) ranging from $A$ dominates (blue), 
equal proportions (green), to $B$ dominates (red). Increases in equilibrium fractions due to heterogeneity are shown in blue shades (right column) and decreases in shades of red. The intensity of the colour indicates the strength of the effect. Accumulated payoffs in heterogenous populations shift the equilibrium in support of the more efficient strategy $A$ except for harmony games where $A$ dominates in any case (bottom right quadrant). Conversely, for averaged payoffs the support of strategy $A$ is much weaker and even detrimental for $T<1+S$. Parameters: initial configuration is a random distribution of equal proportions of strategies $A$ and $B$; each simulation run follows $1.6\cdot10^7$ updates and the equilibrium frequency of $A$ is averaged over the last $2.5\cdot10^6$ updates; results are averaged over $500$ independent runs; for scale-free networks the network is regenerated every $50$ runs. No mutations occured during the simulation run.}
\end{figure}

Here we focus on a related imitation process where an individual $i$ is chosen at random to reassess its strategy by comparing its performance to a randomly chosen neighbour $j$. Individual $i$ then imitates the strategy of $j$ with probability
\begin{align}
\label{eq:imitation}
\frac12 + \frac12\frac{f_j - f_i}{f_j +f_i},
\end{align}
where $f_j$ and $f_i$ are the fitnesses of $i$ and $j$. This variant is convenient as it includes an appropriate normalization factor and hence works regardless of how the fitnesses are calculated. In particular, for exponential payoff-to-fitness mapping (see \eq{fecund}) the imitation rule, \eq{imitation}, recovers the Fermi-update \cite{szabo:PRE:1998}:
\begin{subequations}
\begin{align}
\label{eq:fermi}
&\frac1{1+e^{-\delta(\pi_j-\pi_i)}},&&\text{\footnotesize\emph{\ accumulated}}&\\
&\dfrac1{ 1+e^{-\delta\left(\frac{\pi_j}{n_j}-\frac{\pi_i}{n_i}\right)}},&&\text{\footnotesize\emph{\ averaged}}&
\end{align}
\end{subequations}
For a comparison between averaged and accumulated payoffs in homogenous and heterogenous populations, see \fig{avgaccu}.

On a microscopic level averaging or accumulating payoffs in heterogenous populations turns out to have important biological implications: when averaging payoffs, individuals play different games depending on their location on the graph, whereas for accumulated payoffs everyone plays the same game but at different rates -- again based on the individuals' locations. These intriguing differences are illustrated and discussed for the simplest heterogenous structure, the star graph. First we develop a framework that aids in analyzing an evolutionary process in heterogeneous, graph-structured populations. 

\section*{Criteria for Evolutionary Success}
\label{section:criteria}
In order to determine the evolutionary success of a strategic type in a finite population we consider three fixation probabilities: $\rho_A, \rho_B$ and $\rho_0$. The first, $\rho_A$, indicates the probability that a single $A$ type in an otherwise $B$ population goes on to supplant all $B$s, while the second, $\rho_B$, refers to the probability of the converse process where a single $B$ type takes over a population of $A$ types. These fixation probabilities are important whenever mutations can arise in the population during reproduction or through errors in imitating the strategies of others. The last probability, $\rho_0$, denotes the fixation probability of the neutral process, which is defined as the dynamic in a population with vanishing selection, $\delta=0$. In such a case the game payoffs do not matter and everyone has the same fitness. Based on these fixation probabilities two distinct and complementary criteria are traditionally used to measure evolutionary success \cite{nowak:Nature:2004,
zukewich:PlosOne:2013}:
\begin{enumerate}[(i)]
\item Type $A$ is said to have an evolutionary \emph{advantage} or is \emph{favoured} if
\begin{align}
\label{eq:aadv0}
\rho_A>\rho_B
\end{align}
holds. If mutations, or errors in imitation, are rare the mutant has disappeared or taken over the entire population before the next mutation occurs. We can then view the population dynamic as an embedded Markov chain transitioning between two states: all-$A$ and all-$B$. Denote the proportion of time spent in the state all-$A$ (respectively, all-$B$) by $T_A$ (resp. $T_B$). Together, $T_A$ and $T_B$ are known as the \emph{stationary distribution} of the Markov chain and satisfy the balance equation
\begin{eqnarray}
\label{eq:balanceeq}
T_B \mu_A \rho_A = T_A \mu_B\rho_B,
\end{eqnarray}
where $\mu_A$ ($\mu_B$) is the probability an $A$ ($B$) appears in the all-$B$ (all-$A$) population. For homogeneous populations, or if mutations are not tied to reproduction or imitation events, $\mu_A = \mu_B$ and so Eq. (\ref{eq:balanceeq}) reads
\begin{eqnarray}
T_B \rho_A = T_A\rho_B.
\end{eqnarray}
Hence, if $\rho_A>\rho_B$ then $T_A > T_B$, which captures the notion of $A$ having an advantage over $B$. If the inequality, \eq{aadv0}, is reversed then type $B$ has the advantage.

\item Type $A$ is a \emph{beneficial} mutation if
\begin{subequations}
\label{eq:bene0}
\begin{align}
\label{eq:abene}
\rho_A & >\rho_0\\
\intertext{holds. Similarly, if}
\label{eq:bbene}
\rho_B & >\rho_0
\end{align}
\end{subequations}
holds, the $B$ type is a beneficial mutation. Note that, in general, \eqs{abene}{bbene} are not mutually exclusive. $A$ and $B$ types may simultaneously be advantageous mutants -- in co-existence games, $S>0,T>1$, such as the snowdrift game -- or both disadvantageous -- in coordination games, $S<0,T<1$, such as the stag-hunt game.However, for payoff matrices that satisfy equal-gains-from-switching, such as Table 2, $\rho_A>\rho_0$ implies $\rho_B<\rho_0$ and vice versa in unstructured populations or for weak selection \cite{taylor:JTB:2007}.
\end{enumerate}
The above conditions (12) and (15) are based on the implicit assumption of homogenous populations or averaged payoffs and randomly placed mutants. In the present context of heterogenous populations and with mutants explicitly arising through errors in reproduction or imitation, both conditions require further scrutiny and appropriate adjustments.

The first condition implicitly assumes that an $A$ mutant appears in a monomorphic $B$ population with the same probability as a $B$ mutant in a monomorphic $A$ population. However, in heterogenous populations with accumulated payoffs this is not necessarily the case because even in monomorphic states hubs may have a higher fitness and hence are more readily imitated, or reproduce more frequently, than low degree vertices. This can result in a bias of the rates $\mu_A, \mu_B$ at which $A$ and $B$ mutants arise. Thus, the condition for evolutionary advantage, \eq{aadv0}, must read
\begin{align}
\label{eq:aadv}
\mu_A\rho_A>\mu_B\rho_B.
\end{align}
In general, $\mu_A$ and $\mu_B$ depend on the population structure as well as the payoffs and their accounting. The star structure serves as an illustrative example in the next section. 

Similarly, the second condition also needs to be made more explicit. In general, to determine whether a mutation is beneficial its fixation probability should exceed the probability that in the corresponding monomorphic population one particular individual eventually establishes as the common ancestor of the entire population. We denote these monomorphic fixation probabilities by $\rho_{AA}$, and $\rho_{BB}$, respectively. Thus, the second condition, \eq{bene0}, should be interpreted as
\begin{subequations}
\label{eq:bene}
\begin{align}
\rho_A & >\rho_{BB}\\
\rho_B & >\rho_{AA},
\end{align}
\end{subequations}
i.e. that the fixation probability of a single $A$ (or $B$) mutant in a $B$ ($A$) population exceeds that of one $B$ ($A$) individual turning into the common ancestor of the entire population. 

If mutations occur during an updating event, then in heterogeneous populations mutants occur more frequently in some vertices than in others. For our imitation process, high degree vertices serve more often as models than low degree vertices and hence the mutation is likely to occur in neighbours of high degree vertices. Note that this is different from placing a mutant on a vertex chosen uniformly at random from all vertices \cite{hadjichrysanthou:DGA:2011}. A randomly placed neutral mutant fixates, on average, with a probability corresponding to the inverse of the population size. This is not necessarily the case if neutral mutants arise in reproductive events or errors in imitating or adopting other strategies. In fact, the distinction between $\rho_0, \rho_{AA}$ and $\rho_{BB}$ is only required on heterogenous graphs with accumulated payoffs and non-random locations of mutants. In all other situations the (average) monomorphic fixation probabilities are the same and equal to 
$\rho_0=1/N$, where $N$ is the population size.

In summary, due to the fitness differences in a monomorphic $A$ population with accumulated payoffs the turnover is accelerated and more strategy updates take place and hence more errors occur than in the corresponding monomorphic $B$ population. This means that, on average, mutant $B$s more frequently attempt to invade an $A$ population than vice versa. 

\subsection*{\label{sect:star}The Star Graph}
The star graph represents the simplest, highly heterogenous structure. A star graph of size $N+1$ consist of a central vertex, the hub, which is connected to all $N$ leaf vertices. On the star graph the range of degrees is maximal -- the hub has degree $N$ and all leaves have degree one.

 In order to illustrate the differences arising from accumulating and averaging payoffs, consider a situation where each individual initiated, on average, one interaction. Thus, the hub has $N+1$ interactions while the leaves have only $1+1/N$. Assume that $i$ vertices are of type $A$ and $N-i$ of type $B$. The payoff to a hub of type $A$ is then $(i+(N-i)S)(1+1/N)$ for accumulated payoffs and $(i+(N-i)S)/N$ if payoffs are averaged. In contrast, the payoff of an $A$ leaf is $1+1/N$ (accumulated) and $1$ (averaged). From each $A$ leaf the hub gains $1+1/N$ for accumulated payoffs, which is the same as the gain for the $A$ leaf. However, for averaged payoffs, the hub only gains $1/N$ from each $A$ leaf but each $A$ leaf still gains $1$ from the interaction with the hub. Thus, $A$-$A$-interactions are more profitable for vertices with a low degree and the payoff gets discounted for vertices with larger degrees. Although potential losses against $B$ leaves also get discounted: $T$ for $B$ leaves versus $S/N$ for 
an $A$ hub for averaged payoffs as opposed to $T(1+1/N)$ for $B$ leaves versus $S(1+1/N)$ for an $A$ hub for accumulated payoffs. For $A$ types it is less attractive to interact with $B$ types whenever $S<1$ and hence applies to all generalized social dilemmas \cite{hauert:JTB:2006a}.

\begin{figure}[tbp]
\centerline{\epsfig{file=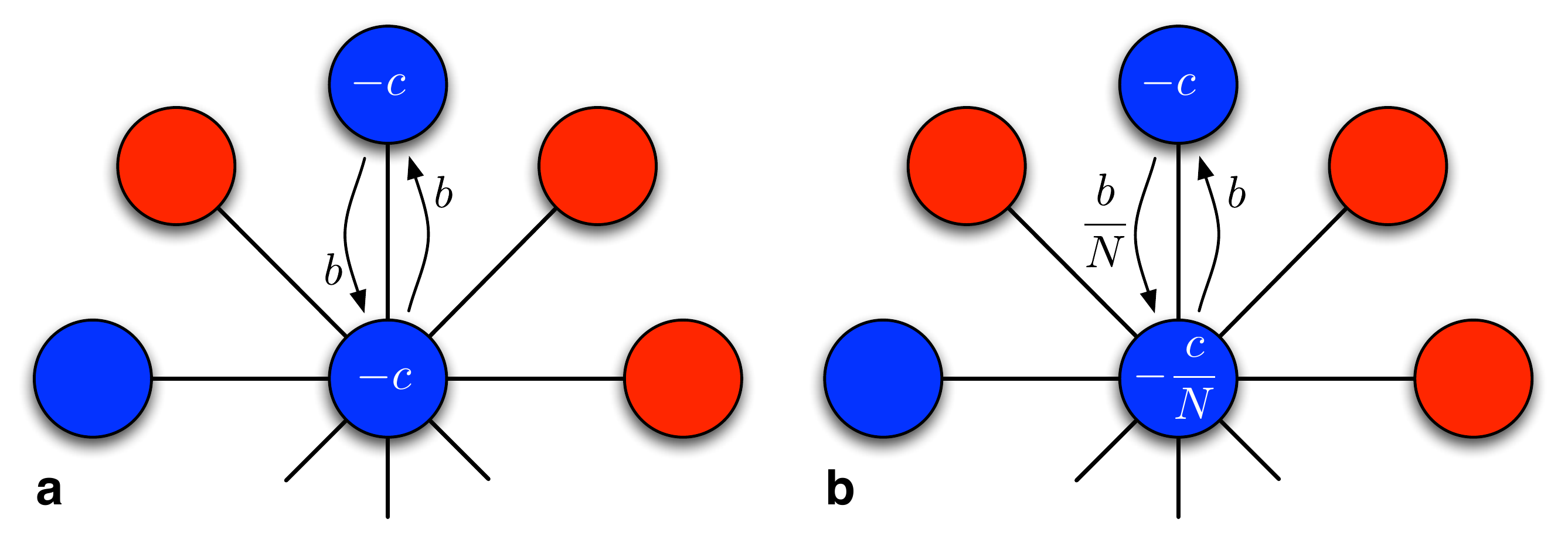,width=0.9\textwidth}}
\caption{A star graph has the hub in the centre surrounded by $N$ leaf vertices. Using the matrix in Table 2, an $A$ type individual (blue) on the hub provides a benefit $b$ to each leaf, regardless of whether the payoffs are \textbf{\textsf a} accumulated or \textbf{\textsf b} averaged. For each interaction, the costs to the hub amount to $c$ in the accumulated case whereas only $c/N$ in the averaged case. Conversely, the costs to a type $A$ leaf are always $c$ and it provides a benefit $b$ to the hub if payoffs are accumulated whereas only $b/N$ when averaged. Hence for averaged payoffs an $A$ type hub provides a benefit to each leaf at a fraction of the costs while $A$ type leaves provide a fraction of the benefits to the hub. This means that the leaves and the hub are playing different games. More specifically, the cost-to-benefit ratio of $A$ leaves is $N c/b$ while it is $c/(N b)$ for an $A$ hub. For most of the population (the leaves), this ratio is much larger than for accumulated payoffs where 
the cost-to-benefit ratio is $c/b$. As a consequence cooperation is much more challenging if payoffs are averaged rather than accumulated.}
\label{fig:star}
\end{figure}

Similarly, the payoffs to a type $B$ hub are $i\, T(1+1/N)$ (accumulated) and $i\,T/N$ (averaged) versus $0$ for $B$ leaves (accumulated and averaged) or $S(1+1/N)$ (accumulated) and $S$ (averaged) for $A$ leaves. In $B$-$B$-interactions both players get zero, regardless of the aggregation of payoffs, which is a consequence of our particular scaling of the payoff matrix in Table 1. Hence there is no discrimination between vertices of different degrees. An illustration of the differences arising from payoff accounting for the simpler and more intuitive case of the prisoner's dilemma in terms of costs and benefits (see Table 2), is given in \fig{star}.
In particular, on star graphs or, more generally, on scale-free networks, averaged payoffs result in higher and hence less favourable cost-to-benefit ratios for most individuals in the population, those with the lower degree vertices. Naturally these differences are also reflected in the evolutionary dynamics. We demonstrate this through the fixation probabilities of a single $A$ ($B$) type in a population of $B$ ($A$) types. 

Let us first consider the fixation probability of a single $A$ type, $\rho_A$. Because of the heterogenous population structure, $\rho_A$ depends on the location of the initial $A$ -- for a star graph, whether the $A$ originated in the hub or one of the leaves. We denote the two fixation probabilities by $\rho_{A|H}$ and $\rho_{A|L}$, respectively. With probability $N/(N+1)$ one of the leaves is chosen to update its strategy and the hub with probability $1/(N+1)$. For averaged payoffs the fitnesses of everyone is the same in a monomorphic $B$ population and hence the hub is equally likely to adopt the strategy of a leaf, and make a mistake with probability $\mu \ll 1$, as are leaves that are adopting the hubs strategy. Hence the average fixation probability is given by 
\begin{align}
\label{eq:rhoAstaravg}
\bar\rho_A = \frac N{N+1}\rho_{A|L} + \frac1{N+1}\rho_{A|H}.
\end{align}
In contrast, for accumulated payoffs even in a homogenous population the hub does not necessarily have the same payoffs as the leaves because of the larger number of interactions. However, for our payoff matrix in Table 1, this does not matter for homogenous $B$ populations as all $B$-$B$-interactions yield a payoff of zero. Consequently, \eq{rhoAstaravg} equally holds for averaged and accumulated payoffs and, incidentally, this is also the average fixation probability for a randomly placed $A$ mutant.

Similarly, we are interested in the average fixation probability, $\bar\rho_B$, of a single $B$ type in an otherwise homogenous $A$ population. Again we first need to determine with what probability the $B$ mutant arises in a leaf or in the hub. Interestingly, and in contrast to $\bar\rho_A$, this now depends on the accounting of payoffs. If payoffs are averaged then all individuals have the same payoff and, in analogy to \eq{rhoAstaravg}, we obtain
\begin{align}
\label{eq:rhoBstaravg}
\bar\rho_B^\text{\ avg} = \frac N{N+1}\rho_{B|L} + \frac1{N+1}\rho_{B|H}.
\end{align}
However, for accumulated payoffs, the hub achieves a payoff of $N+1$ as compared to an average payoff of merely $1+1/N$ for the leaves. In order to determine the average fixation probability of a single $B$ type, $\bar\rho^\text{\ accu}_B$, we first consider the case where the mutant arises on a leaf. With probability $N/(N+1)$ a leaf is selected to update its strategy and adopts the hub's strategy with probability $1/(1+\exp(-\delta(N-1/N)))$ (c.f. \eq{fermi}). If the leaf adopts the strategy it makes an error with a small probability and instead of copying the $A$ strategy, the leaf becomes of type $B$. Similarly, the hub reassesses its strategy with probability $1/(N+1)$ and switches to the leafs strategy with probability $1/(1+\exp(\delta(N-1/N)))$, which may then give rise to an $A$ type in the hub with a small probability. Based on these probabilities we can now determine the proportion of mutants that occur in the leaves and the hub, respectively. For the leaves we get
\begin{align}
\frac{\frac N{N+1}\frac1{1+e^{-\delta\left(N-\frac1N\right)}}}{\frac N{N+1}\frac1{1+e^{-\delta\left(N-\frac1N\right)}}+\frac1{N+1}\frac1{1+e^{\delta\left(N-\frac1N\right)}}} = \frac N{N+e^{-\delta\left(N-\frac1N\right)}}\notag
\intertext{and similarly for the hub}
\frac{\frac1{N+1}\frac1{1+e^{\delta\left(N-\frac1N\right)}}}{\frac N{N+1}\frac1{1+e^{-\delta\left(N-\frac1N\right)}}+\frac1{N+1}\frac1{1+e^{\delta\left(N-\frac1N\right)}}} = \frac1{1+N\,e^{\delta\left(N-\frac1N\right)}}.\notag
\end{align}
Thus, the average fixation probability of a single $B$ mutant is
\begin{align}
\label{eq:rhoBstaraccu}
\bar\rho_B^\text{\ accu} = \frac N{N+e^{-\delta\left(N-\frac1N\right)}}\ \rho_{B|L} + \frac 1{1+Ne^{\delta\left(N-\frac1N\right)}}\ \rho_{B|H}.
\end{align}

In the weak selection limit, $\delta\ll1$ (or, more precisely,  $\delta N\ll1$), \eq{rhoBstaraccu} takes on the same form as for averaged payoffs, \eq{rhoBstaravg}. Conversely, for large populations, $\delta N\gg1$, mutants almost surely arise in leaves and hence $\bar\rho_B^\text{\ accu}\approx\rho_{B|L}$. Note that this is a good approximation as for $N=100$ and $\delta=0.1$ the probability that the mutant arises in the hub is already less than $10^{-6}$.

In order to determine the evolutionary advantage of $A$ and $B$ types we still need to determine the rates $\mu_A, \mu_B$ at which $A$ and $B$ mutants arise in monomorphic $B$ and $A$ populations, respectively. If payoffs are averaged all individuals in the population have the same fitness and hence with probability $1/2$ the focal individual imitates its neighbour (c.f. \eq{fermi}) and with a small probability $\mu$ an error (or mutation) occurs. This holds for monomorphic populations of either type and hence $\mu_A=\mu_B$. For accumulated payoffs the same argument holds for monomorphic $B$ populations where all individuals have zero payoff. Consequently, $A$ mutants arise at a rate $\mu_A=1/2\mu$. In contrast, in a monomorphic $A$ population the hub has a much higher fitness and leaves will almost surely imitate the hub (whereas the hub almost surely will not imitate a leaf):
\begin{align}
\mu_B &= \left(\frac N{N+1}\frac1{1+e^{-\delta\left(N-\frac1N\right)}}+\frac1{N+1}\frac1{1+e^{\delta\left(N-\frac1N\right)}}\right) \mu.
\end{align}
For large $N$ every update essentially results in one of the leaves imitating the hub, so that $\mu_B\approx \mu$.

\Eqss{rhoAstaravg}{rhoBstaraccu} yield the conditions under which type $A$ or $B$ has an evolutionary advantage. For star graphs, the fixation probabilities, $\rho_A$ and $\rho_B$, can be derived based on the transition probabilities to increase or decrease the number of mutants by one and hence the results can be easily applied to any update rule \cite{hadjichrysanthou:DGA:2011}. For the imitation dynamics $A$ types are favoured under weak selection if and only if
\begin{subequations}
\begin{align}
\label{eq:ABstaraveraged}
\bar\rho_A^\text{\ avg} &> \bar\rho_B^\text{\ avg} & \Longleftrightarrow && \frac{N-1}{2N} & >T-S && \text{\emph{averaged}}\\
\mu_A\bar\rho_A^\text{\ accu} &> \mu_B\bar\rho_B^\text{\ accu} & \Longleftrightarrow && \frac{2N(N+1)}{N^2+4N-1} & >T-S 
 && \text{\emph{accumulated}}
\end{align}
and in the limit of infinite populations, $N\to\infty$, the conditions reduce to
\begin{align}
\bar\rho_A^\text{\ avg} &> \bar\rho_B^\text{\ avg} & \Longleftrightarrow && \frac12 & >T-S && \text{\emph{averaged}}\\
\label{eq:ABstaraccumulatedNinfty}
\mu_A \bar\rho_A^\text{\ accu} &> \mu_B \bar\rho_B^\text{\ accu} & \Longleftrightarrow && 2 &> T-S && \text{\emph{accumulated}}
\end{align}
\end{subequations}
A detailed derivation of the different fixation probabilities is provided in the Materials and Methods section. 

In order to determine whether a mutant is favoured or not (see \eq{bene}), we first need to determine the fixation probabilities $\rho_{AA}$ and $\rho_{BB}$. Naturally, those fixation probabilities again depend on whether the ancestor is located in the hub or one of the leaves. Let us first consider a monomorphic $B$ population. The fixation probability of a $B$ located in the hub, $\rho_{BB|H}$, or in one particular leaf, $\rho_{BB|L}$, can be derived from the fixation probabilities $\rho_{B|H}$ and $\rho_{B|L}$ by setting $f_i=1$ (see Materials and Methods), which yields
\begin{subequations}
\begin{align}
\rho_{BB|H} & = \frac12\\
\rho_{BB|L} & = \frac1{2N}.
\end{align}
\end{subequations}
Intuitively, the hub individual becomes the common ancestor with probability $1/2$ because any leaf individual updates its strategy to the hub's with a probability of $1/2$ and the hub keeps its strategy also with probability of $1/2$ but both probabilities are independent of the size of the population. Conversely, a leaf individual must first be imitated by the hub, which is $1/N$ times less likely than the reverse. On average we then obtain (insert into \eq{rhoAstaravg}):
\begin{align}
\label{eq:rhobb}
\bar\rho_{BB} & = \frac1{N+1}.
\end{align}
Note that in a monomorphic $B$ population the payoffs are zero regardless of the selection strength, $\delta$, location (hub and leaves) or the payoff accounting. Again, this is a consequence of our particular choice of payoff matrix (Table 1), and thus, \eq{rhobb} holds for both averaged as well as accumulated payoffs and is, in fact, the same as the neutral fixation probability $\rho_0$.

\begin{figure}[tbp]
\centerline{\epsfig{file=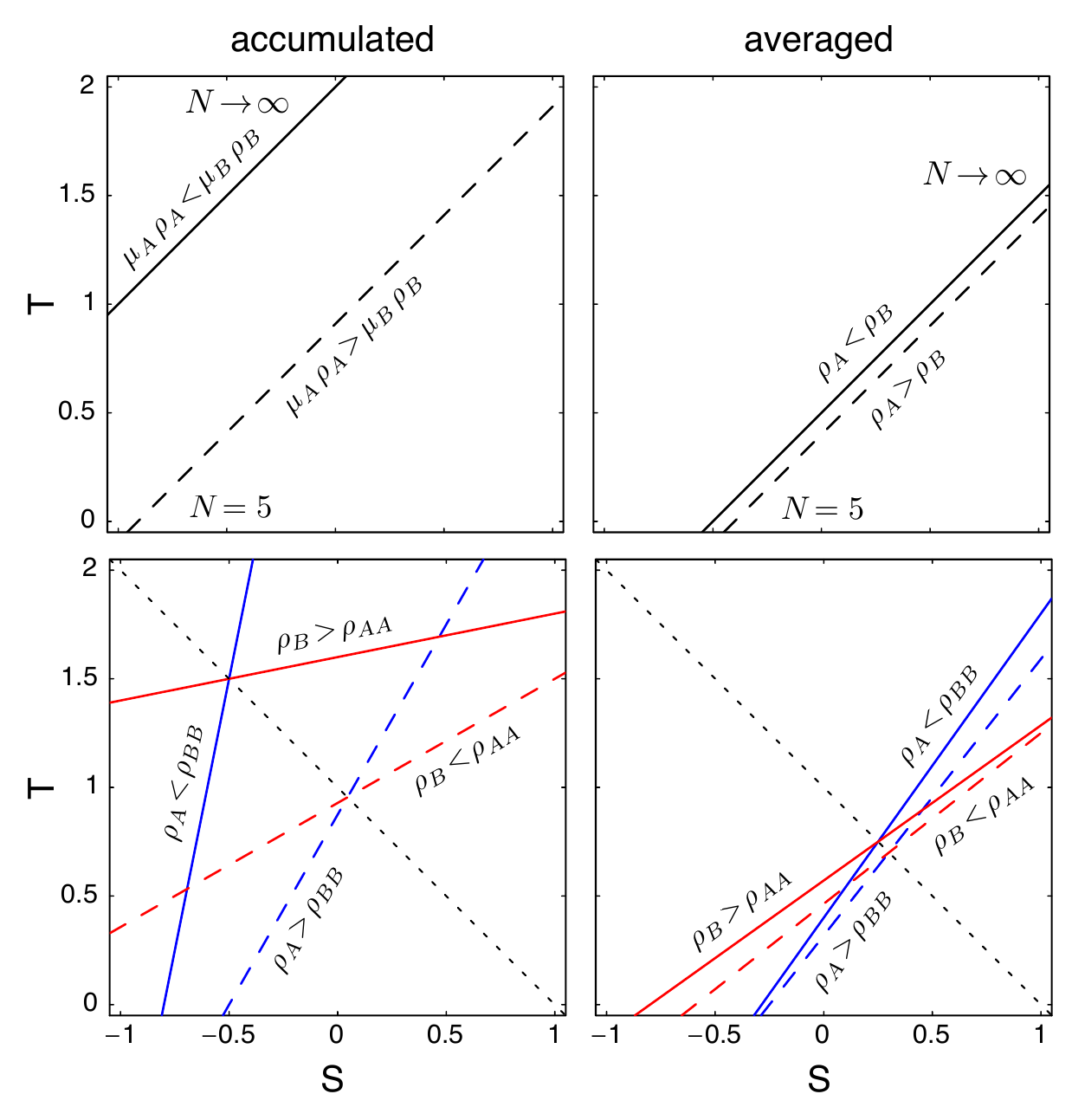,width=0.7\textwidth}}
\caption{\label{fig:rhoAB}Criteria for evolutionary success on the star graph for accumulated (left column) and averaged (right column) payoffs for weak selection, $\delta\ll1$. The range for which $A$ is advantageous (top row, c.f. \eq{aadv}) depends on the population size, $N$, and is shown in the limit $N\to\infty$ (solid line) and for $N=5$ (dashed line). Below the respective lines $A$ is favoured. Similarly, the range for which $A$ and $B$ mutants are beneficial (c.f. \eq{bene}) also depends on $N$. $B$ mutants are beneficial above the red lines, while $A$ mutants are beneficial below the blue lines (solid for $N\to\infty$; dashed for $N=5$). Additive games (or equal-gains-from-switching) satisfy $S+T=1$ (dotted line).}
\end{figure}

Let us now turn to the monomorphic $A$ population and determine $\rho_{AA}$. If $\delta=0$ then everything is the same as in the monomorphic $B$ population above and $\bar\rho_{AA}=1/(N+1)$. However, for any non-zero selection, $\delta>0$, the situation becomes more interesting. If payoffs are averaged, all individuals have the same (non-zero) payoffs and a mutant is equally likely to appear in the hub as any particular leaf (c.f. \eq{rhoBstaravg}) and hence $\bar\rho_{AA}=1/(N+1)$ still holds. However, if payoffs are accumulated the hub has a higher fitness. The fixation probabilities that an $A$ on the hub or one of the leaves becomes the common ancestor are $\rho_{AA|H}$ and $\rho_{AA|L}$ (see Materials and Methods) and, on average we obtain
\begin{align}
\bar\rho_{AA}^\text{\ accu} & = \frac1{N+1} -\frac12\left(\frac{N-1}{N+1}\right)^2 \delta + O\left(\delta^2\right).
\end{align}
Now we are able to derive the conditions under which an $A$ and/or $B$ mutant is beneficial, c.f. \eq{bene}:
\begin{subequations} 
\begin{align}
\label{eq:AABB1}
\bar\rho_A^\text{\ avg} &> \bar\rho_{BB}^\text{\ avg} & \Longleftrightarrow && (4N^2-3N-1) + (14N^2-3N+1)S &> (10N^2+3N-1)T \\
\bar\rho_B^\text{\ avg} &> \bar\rho_{AA}^\text{\ avg} & \Longleftrightarrow && (8N^2-9N+1) + (10N^2+3N-1)S &< (14N^2-3N+1)T 
\intertext{for averaged payoffs and, for accumulated payoffs,}
\bar\rho_A^\text{\ accu} &> \bar\rho_{BB}^\text{\ accu} & \Longleftrightarrow && (4N^2-3N-1) + (5N^2+9N-2)S &> (N^2+15N-4)T \\
\bar\rho_B^\text{\ accu} &> \bar\rho_{AA}^\text{\ accu} & \Longleftrightarrow && (8N^2-9N+1) + (N^2+15N-4)S &< (5N^2+9N-2)T.
\label{eq:AABB4}
\end{align}
\end{subequations}
The parameter region which delimits the region of evolutionary success of $A$ and $B$ types is illustrated in \fig{rhoAB}. 

We can analyze Eqs. (\ref{eq:ABstaraveraged}) - (\ref{eq:ABstaraccumulatedNinfty}) and (\ref{eq:AABB1}) - (\ref{eq:AABB4}) in terms of the additive prisoner's dilemma game by substituting $S=-c/(b-c)$ and $T=b/(b-c)$. For simplicity, we restrict attention to the case $N\to \infty$ and since in the additive prisoner's dilemma game a strategy is favoured if and only if it beneficial we need only consider Eqs. (\ref{eq:ABstaraveraged}) - (\ref{eq:ABstaraccumulatedNinfty}). We have
\begin{subequations}
\begin{align}
\label{eq:bcaveonstars}
\bar\rho_A^\text{\ avg} &> \bar\rho_B^\text{\ avg} & \Longleftrightarrow && \frac{b}{c} & < -3 && \text{\emph{averaged}}\\
\label{eq:bckonstars}
\mu_A \bar\rho_A^\text{\ accu} &> \mu_B \bar\rho_B^\text{\ accu} & \Longleftrightarrow && \frac{b}{c} &> 3 && \text{\emph{accumulated}}
\end{align}
\end{subequations}
If we suppose $b,c >0$, then Eq. (\ref{eq:bcaveonstars}) is never satisfied. That is, averaging rather than accumulating the payoffs altogether removes the ability of the star graph to support cooperation. 

Note that for additive, or equal-gains-from-switching, games (games that satisfy $S+T=1$) and for weak selection the condition $\rho_A>\rho_{BB}$ implies both $\rho_B<\rho_{AA}$ and $\rho_{BB}=\rho_{AA}=1/(N+1)$, regardless of the accounting of payoffs. This extends results obtained for homogenous populations \cite{ohtsuki:Nature:2006,taylor:Nature:2007}. 

\section*{Stochastic Interactions \& Updates}
As we have seen, when payoffs are averaged, members of a heterogeneous population are possibly playing different games, while if they are accumulated, all individuals play the same game. Therefore, only accumulating payoffs allows for meaningful comparissons of different heterogeneous population structures. A common simplifying assumption is that each individuals interacts once with all its neighbours, see \fig{avgaccu}. For heterogeneous populations this assumption means that those individuals residing on higher-degree vertices are interacting with their neighbours at a higher rate than those on lower-degree vertices. This leads to a separation of time scales, where interactions occur on a much faster time scale than strategy updates.

Realistically, all social interactions require a finite amount of time and hence the number of interactions per unit time is limited. This constraint already affects the evolutionary process in unstructured populations \cite{woelfing:JTB:2009} but becomes particularly important in heterogenous networks where, for example, in scale-free networks some vertices entertain neighbourhood sizes that are orders of magnitude larger than that of other vertices. For those hubs it may not be possible to engage in interactions with all neighbours between subsequent updates of their strategy or the strategies of one of their neighbours. In order to investigate this we need to abandon the separation of the timescales for interactions and strategy updates.

\begin{figure}[tbp]
\centerline{\epsfig{file=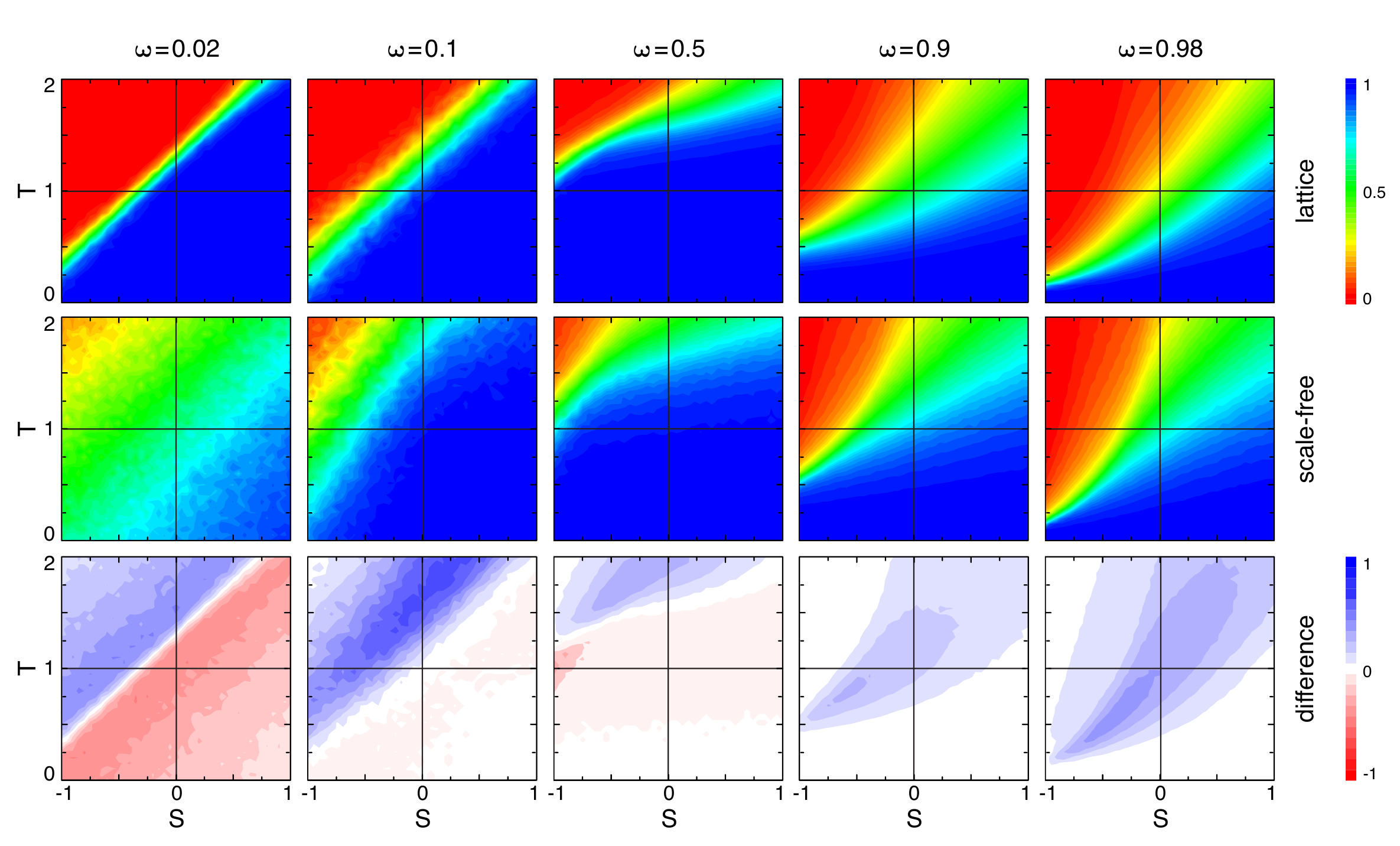,width=\textwidth}}
\caption{\label{fig:rates}Average fraction of strategy $A$ for different ratios between interactions and strategy updates in homogenous (top row) and heterogeneous (middle row) populations and the difference between them (bottom row) as a function of the game parameters $S$ and $T$ (c.f. \fig{avgaccu}). Interactions occur with probability $\omega$ and strategy updates with $1-\omega$. For example, for $\omega=0.9$ each individual has, on average, initiated $9$ interactions between strategy updates but only an average of $1/9$ interactions for $\omega=0.1$. For small $\omega$ effects of heterogenous population structures have little chance to manifest themselves and the results are closer to those for averaged payoffs (c.f. \fig{avgaccu}). In contrast, for large $\omega$ heterogeneity plays an important role: for scale-free networks it is guided by the structural heterogeneity whereas in homogenous populations another form of heterogeneity spontaneously emerges in the number of interactions. Even on lattices, 
stochastic differences in the number of interactions get amplified by the dynamics because an increased number of interactions reduces the chances that an individual updates its strategy (c.f. \fig{idistr}). As a consequence the results for lattices and scale-free networks become increasingly similar but scale-free networks keep promoting $A$ types to a greater extend. Parameters and averaging technique are as in the caption to \fig{avgaccu}.
}
\end{figure}

A unified time scale on which interactions and strategy updates occur can be introduced as a stochastic process where a randomly chosen individual $i$ initiates an interaction with probability $\omega$ with a random neighbour $j$ and reassesses its strategy with probability $1-\omega$ by comparing its payoff to that of a random neighbour according to \eq{imitation}. Interactions alter the payoffs $\pi_i, \pi_j$ of both individuals (and hence their fitnesses, $f_i,f_j$, see \eq{accufecund}) according to the game matrix in Table 1. If individual $i$ adopts the strategy of its neighbour, then its payoff (and interaction count) is reset to zero, $\pi_i=0$, regardless of whether the imitation had resulted in an actual change of strategy. Simulation results for various $\omega$ are shown in \fig{rates}.

For small $\omega\ll1$ few interactions occur between strategy updates and in the limit $\omega\to0$ neutral evolution is recovered because no interactions occur. Conversely, in the limit $\omega\to1$ many interactions occur between strategy updates, which allows individuals to garner large payoffs as well as build up large payoff differences. The average number of interactions initiated by any individual between subsequent reassessments of the strategy is $\omega/(1-\omega)$, the relative ratio of the time scales of game interactions versus strategy updates. However, the distribution of the number of interactions is biased: individuals with a large number of interactions tend to score high payoffs and hence are less likely to imitate a neighbours' strategy, which in turn results in a further increase of interactions. On heterogenous graphs and scale-free networks, in particular, this bias is built-in by the underlying structure because highly connected hubs engage, on average, in a much larger number of 
interactions than vertices with few neighbours. Moreover, hubs are more likely to serve as models when neighbours are reassessing their strategy -- simply because hubs have many neighbours. Thus, hubs are not only more resilient to change but also have a stronger influence on their neighbourhood. When $\omega/(1-\omega)$ this ratio begins to get large, interactions dominate strategy updates and the resulting game dynamics on heterogeneous and homogeneous graphs becomes indistinguishable.

\begin{figure}[tbp]
\centerline{\epsfig{file=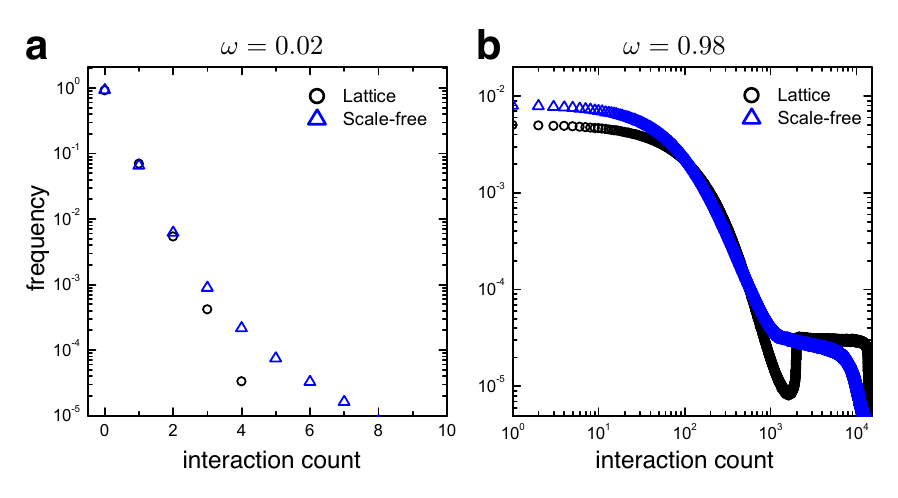,width=0.8\textwidth}}
\caption{\label{fig:idistr}Distributions of the number of interactions on lattices (black {\large$\circ$}) and scale-free networks (blue {\small$\triangle$}) with \textbf{\textsf{a}} few interactions between updates ($\omega=0.02$ or, on average, $\approx0.02$ interactions) and \textbf{\textsf{b}} many interactions between updates ($\omega=0.98$ or, on average, $49$ interactions). For small $\omega$ the heterogeneity of scale-free networks results in a pronounced tail at higher numbers of interactions compared to the approximately exponential distribution for lattices. This tail is responsible for the reduction of cooperation in scale-free networks observed in \fig{rates}: as interactions dominate, some vertices almost never update their strategies. This "static network" emerges in both lattices and scale-free graphs and prevents the complete proliferation of the rare strategy. Nevertheless, most of the individuals in the population experience essentially the same number of interactions. The distributions 
look different for large $\omega$ but the main difference remains that scale-free networks produce a more pronounced tail. More importantly, however, for most of the population the distributions are actually very similar and hence the heterogeneities very similar. On lattices, the skewed distribution is caused by stochastic variations and the positive feedback between the number of interactions and the resilience to changing strategy.
}
\end{figure}

Interestingly, a similar bias in interaction numbers spontaneously emerges on homogenous graphs, lattices in particular. Since all vertices have the same number of neighbours, no vertices are predisposed to achieve more interactions than others but some inequalities in interaction numbers occur simply based on stochastic fluctuations. As above, those vertices that happen to engage in more interactions tend to have higher payoffs and hence are less likely to imitate their neighbours and keep aggregating payoffs. This positive feedback between interaction count and resilience to change spontaneously introduces another form of heterogeneity, which becomes increasingly pronounced for larger $\omega$. In fact, for large $\omega$ it rivals the structurally imposed heterogeneity of scale-free networks, see \fig{idistr}.

Regardless of the structure, the positive feedback between payoff aggregation and the diminishing chances to change strategy (and hence reset payoffs) means that a small set of nodes forms an almost static backdrop of the dynamics and hence has a considerable effect on the evolutionary process. This set is a random selection on homogenous structures and consists of the hubs on heterogenous structures. As a consequence, the initial configuration of the population has long lasting effects on the abundance of strategies.

\begin{figure}[!ht]
\centerline{\epsfig{file=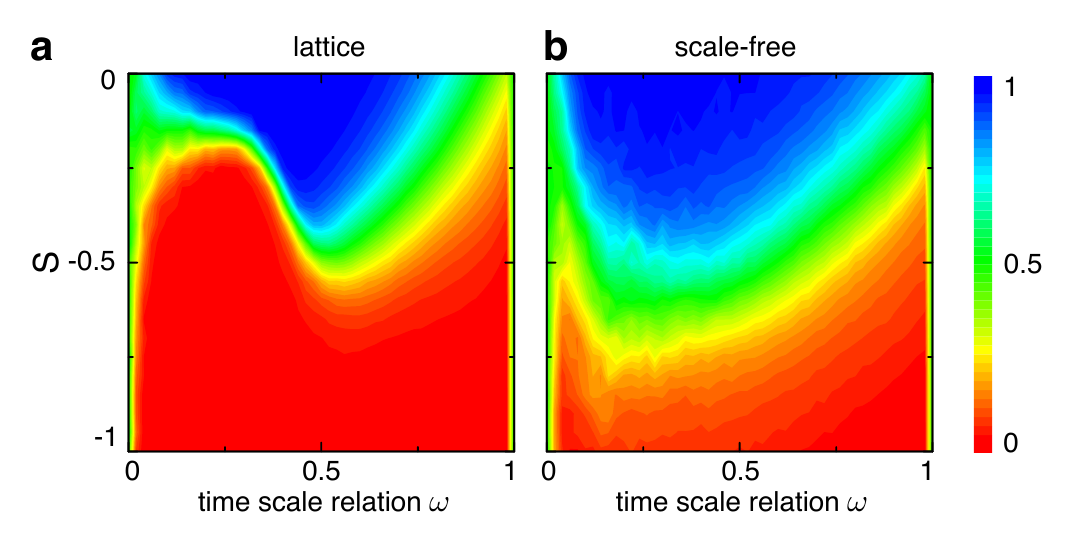,width=0.8\textwidth}}
\caption{\label{fig:omega} Impact of the time scale relation $\omega$ on the equilibrium fractions of cooperators in the $S\omega$-plane for additive prisoner's dilemma games ($T=1-S$): \textbf{\textsf{a}} lattices and \textbf{\textsf{b}} scale-free networks. The limit $\omega\to0$ recovers the neutral process (no interactions) whereas for $\omega\to1$ individuals hardly update their strategies. Thus, in both of the two limiting cases the fraction of cooperators remains at the initial value of $0.5$. For both types of population structures there exists an intermediate $\omega$ that leads to an optimal level of cooperation. On lattices the support for cooperation is strongest if interactions and strategy updates occur at equal rates, $\omega=0.5$, but on scale-free networks more frequent updates than interactions are even more beneficial, $\omega\approx0.25$. Parameters and averaging technique are as in the caption to  \fig{avgaccu}.
}
\end{figure}

A more detailed view on the effects of $\omega$ on the evolutionary process is provided by restricting the attention to the prisoner's dilemma and additive payoffs, c.f. Table 2. This can be accomplished by setting $T=1-S$ with $S<0$. The equilibrium levels of cooperation in the $S\omega$-plane are shown in \fig{omega} for lattices and scale-free networks. 
Altering the relative rates of interactions versus strategy updates has interesting effects on the evolutionary outcome. For lower rates of interaction ($\omega \ll 1$), scale-free networks outperform lattices in their ability to promote cooperation. As interaction rates increase and strategy updates become more rare ($\omega\approx 1$), scale-free networks and lattices become virtiually indistinguishable in their ability to support cooperation. For both lattices and scale-free networks an optimal ratio between strategy updates and interactions exist: for lattices this is roughly $\omega=1/2$, suggesting that lattices support the greatest amount of cooperators when interactions occur at the same rate as strategy updates, whereas for scale-free networks the optimum lies around $\omega\approx0.25$, which suggests that scale-free networks provide the strongest support for cooperation if there are roughly three updates per interaction.

\section*{Discussion}
Evolutionary dynamics in heterogenous populations, scale-free networks in particular, have attracted considerable attention over recent years. Somewhat surprisingly, the underlying microscopic processes and their implications for the macroscopic dynamics and the corresponding biological interpretations have received little attention.

Here we have shown that established criteria to measure success in evolutionary processes make different kinds of implicit assumptions that do not hold in general for heterogenous structures. Instead, for such structures it becomes imperative to reconsider, revise and generalize these criteria, which was done in the Criteria for Evolutionary Success section. If errors arise in imitating the strategic type of other individuals, or mutations occur during reproduction, then mutations are more likely to arise in some locations than in others. For example, on the star graph mutants likely occur in the leaf nodes for birth-death updating and imitation processes but in the hub for death-birth processes. Moreover, in heterogenous populations the fixation probabilities generally depend on the initial location of the mutant and hence even the fixation probability of a neutral mutant may no longer simply be the reciprocal of the population size but rather intricately depend on the population structure.

Another crucial determinant of the evolutionary dynamics in heterogenous populations is the aggregation of payoffs from interactions between individuals. Individuals on vertices with a higher (lower) degree expect to have more (fewer) interactions than on average. Even though the choice between averaging or accumulating payoffs may seem innocuous, it has far reaching consequences. If payoffs are accumulated, some individuals are capable of accruing more payoffs than others strictly by virtue of them having more potential partners. Averaging payoffs removes the ability of hubs to accrue greater payoffs, but simultaneously makes it difficult to compare results for different population structures (e.g. lattices versus scale-free networks) even if their average degrees are the same because the type of game played depends on the location in the graph. Hence, accumulating payoffs seems a more natural choice to compare evolutionary outcomes based on different population structures because it ensures that everyone 
engages in the same game. However, if we assume all interactions are realised then those individuals with more neighbours interact at a much greater rate than those with less. 

In order to investigate the disparity in the number of interactions on the success of strategies on heterogenous graphs we introduced the time-scale parameter $\omega$, which determines the probability that an interaction or a strategy update occurs. When increasing the rate of strategy updates (small $\omega$), heterogeneous graphs are able to support higher levels of cooperation than lattices.  Conversely, increasing the rate of interactions (large $\omega$) results in small differences between lattices and scale-free networks; both support roughly the same levels of cooperation. For imitation processes, individuals with high payoffs are unlikely to change their strategies and hence are likely to keep accumulating more payoffs. On scale-free networks, hubs are predestined to become such high performing individuals but on lattices they spontaneously emerge, triggered by stochastic fluctuation in the interaction count and driven by the positive feedback between increasing payoffs and increasing resilience to 
changing strategies (and hence to resetting payoffs).


For intermediate $\omega$ an optimum increase in the level of cooperation is found: lattices support cooperation most efficiently if a balance is struck between interactions and strategy updates ($\omega\approx0.5$), whereas scale-free networks work most efficiently if slightly more updates occur ($\omega\approx0.25$). For lattices a related observation was reported for noise in the updating process \cite{szabo:PRE:2005}. If the noise is large, updating is random but if it is small the game payoffs become essential. Interestingly, cooperation is most abundant for intermediate levels of noise -- which is similar to having some but not too many interactions between strategy updates.

Previous work has found that heterogeneous graphs support coordination of strategies, where all individuals are inclined to adopt the same strategy, while homogeneous graphs support co-existence \cite{pinheiro:PLOS:2012, pinheiro:NJP:2012}. The time scale parameter $\omega$ introduced in the Stochastic Interactions and Updates section seems to aid in promoting coexistence in both types of graphs, based on the large green region in Figures \ref{fig:avgaccu}, \ref{fig:rates}, and \ref{fig:omega}. Exactly how the time scale parameter $\omega$ promotes coexistence is a topic worthy of further investigation.

Naturally there is no correct way of modelling the updating of the population or the aggregation of payoffs but, as so often, the devil is in the detail and implicit assumptions originating in traditional, homogenous models may be misleading or have unexpected consequences in more general, heterogenous populations.

\section*{Materials and Methods}

In \cite{hadjichrysanthou:DGA:2011}, the authors calculate expressions for the probability that a single mutant fixes on a star graph. These expressions are in terms of state transition probabilities. Denote by $P_{i,j}^{XY}$ the transition probability from a state with $i$ $A$ individuals on the leaves and an $X$ individual on the hub to a state with $j$ $A$ individuals on the leaves and a $Y$ on the hub. With this notation, the fixation probability of a single $A$ on a leaf vertex is
\begin{eqnarray}
\label{eq:broomleaffix}
\rho_{A|L} = \dfrac{P_{0,1}^{AA}}{P_{0,1}^{AA}+P_{1,1}^{AB}}\dfrac{1}{A(1,N)},
\end{eqnarray}
and for a single $A$ on the hub,
\begin{eqnarray}
\label{eq:broomhubfix}
\rho_{A|H} =  \dfrac{P_{1,1}^{BA}}{P_{1,1}^{BA}+P_{1,0}^{BB}}\dfrac{1}{A(1,N)},
\end{eqnarray}
where, in both cases,
\begin{eqnarray}
A(1,N) = \displaystyle 1+\sum_{j=1}^{N-1}\dfrac{P_{j,j}^{AB}}{P_{j,j}^{AB} + P_{j,j+1}^{AA}}\prod_{k=1}^j\dfrac{P_{k,k-1}^{BB}\left( P_{k,k+1}^{AA} + P_{k,k}^{AB}  \right)}{P_{k,k+1}^{AA}\left( P_{k,k-1}^{BB}+P_{k,k}^{BA} \right) }.
\end{eqnarray}
For the imitation process defined by Eq. \ref{eq:imitation} and accumulated payoffs we have
\begin{subequations}
\begin{align}
P_{i,i+1}^{AA} = & \dfrac{N-i}{N+1}\dfrac{e^{\delta(i + (N-i)S)\left(1+1/N\right)}}{e^{\delta(i + (N-i)S)\left(1+1/N\right)} + e^{\delta T\left(1+1/N\right)}}  \\
P_{i,i}^{AB} = & \dfrac{1}{N+1}\dfrac{N-i}{N}\dfrac{e^{\delta T\left(1+1/N\right)}}{e^{\delta(i + (N-i)S)\left(1+1/N\right)} + e^{\delta T\left(1+1/N\right)}} \\
P_{i,i}^{BA} = & \dfrac{1}{N+1}\dfrac{i}{N}\dfrac{e^{\delta S\left(1+1/N\right)}}{e^{\delta iT\left(1+1/N\right)} + e^{\delta S\left(1+1/N\right)}}\\
P_{i,i-1}^{BB} = & \dfrac{i}{N+1}\dfrac{e^{\delta i T\left(1+1/N\right)}}{e^{\delta iT\left(1+1/N\right)} + e^{\delta S\left(1+1/N\right)}}
\end{align}
\end{subequations}
and for averaged payoffs,
\begin{subequations}
\begin{align}
P_{i,i+1}^{AA} = & \dfrac{N-i}{N+1}\dfrac{e^{\delta((i + (N-i)S)/N)}}{\displaystyle e^{\delta((i + (N-i)S)/N)} + e^{\delta T}}  \\
P_{i,i}^{AB} = & \dfrac{1}{N+1}\dfrac{N-i}{N}\dfrac{e^{\delta T}}{e^{\delta((i + (N-i)S)/N)} + e^{\delta T}} \\
P_{i,i}^{BA} = & \dfrac{1}{N+1}\dfrac{i}{N}\dfrac{e^{\delta S}}{e^{\delta(i/N) T} + e^{\delta S}}\\
P_{i,i-1}^{BB} = & \dfrac{i}{N+1}\dfrac{e^{\delta (i/N) T}}{e^{\delta (i/N)T} + e^{\delta S}}.
\end{align}
\end{subequations}
These are incorporated into the Eqs. (\ref{eq:broomleaffix}) and (\ref{eq:broomhubfix}) to yield the fixation probabilities $\rho_{A|L}$ and $\rho_{A|H}$. The fixation probabilities $\rho_{B|L}$ and $\rho_{B|H}$ are obtained in a similar way. The averages $\overline{\rho}_{A,B}^{\text{accu}}$ and $\overline{\rho}_{A,B}^{\text{avg}}$ are then calculated using Eqs. (\ref{eq:rhoAstaravg}), (\ref{eq:rhoBstaravg}), and (\ref{eq:rhoBstaraccu}). Finally, a first-order approximation in $\delta$ is found for the above. 
For example,
\begin{subequations}
\begin{align}
\overline{\rho}_{A}^{\text{accu}} = & \left[\dfrac{1}{N+1}\rho_{A|H} + \dfrac{N}{N+1} \rho_{A|L}\right]_{\delta =0} + \dfrac{d}{d\delta}\left[\dfrac{1}{N+1}\rho_{A|H} + \dfrac{N}{N+1} \rho_{A|L}\right]_{\delta=0}\delta + O(\delta^2 ) \nonumber \\
= & \dfrac{1}{N+1} + \left(\left(\dfrac{1}{N+1}\right)\left. \dfrac{d}{d\delta}\rho_{A|H}\right|_{\delta=0} + \left(\dfrac{N}{N+1}\right)\left. \dfrac{d}{d\delta}\rho_{A|L}\right|_{\delta=0}\right) \delta +O(\delta^2) \nonumber \\
= & \dfrac{1}{N+1} + \dfrac{\delta}{12N(N+1)}\left(\left(5N^2+9N-2\right)S-\left(N^2+15N-4\right)T\right. \nonumber \\
& \left.+\left(4N^2-3N-1\right)\right)+O(\delta^2) 
\intertext{The other fixation probabilities are found in a similar way:}
\overline{\rho}_{B}^{\text{accu}} = & \dfrac{1}{N+1} - \dfrac{\delta}{12N(N+1)^2}\left(\left(N^3+16N^2+11N-4\right)S -\left(5N^3+14N^2+7N-2\right)T \right. \nonumber \\
& \left .+ \left(14N^3-13N^2-2N+1\right) \right) +O(\delta^2) \\ 
\overline{\rho}_{A}^{\text{avg}} = & \dfrac{1}{N+1} + \dfrac{\delta}{12N(N+1)^2}\left(\left(14N^2-3N+1\right)S -\left(10N^2+3N-1\right)T \right. \nonumber \\
& \left .+ \left(4N^2-3N-1\right) \right) +O(\delta^2) \\ 
\overline{\rho}_{B}^{\text{avg}} = & \dfrac{1}{N+1} - \dfrac{\delta}{12N(N+1)^2}\left(\left(10N^2+3N-1\right)S -\left(14N^2-3N+1\right)T \right. \nonumber \\
& \left .+ \left(8N^2-9N+1\right) \right) +O(\delta^2) 
\end{align}
\end{subequations}
Assuming $\delta \ll 1$, and employing the appropriate condition for evolutionary advantage, yields Eqs. (\ref{eq:AABB1}--\ref{eq:AABB4}) in the main text. 

\section*{Acknowledgments}

\bibliography{./ET}

\begin{thebibliography}{10}
\providecommand{\url}[1]{\texttt{#1}}
\providecommand{\urlprefix}{URL }
\expandafter\ifx\csname urlstyle\endcsname\relax
  \providecommand{\doi}[1]{doi:\discretionary{}{}{}#1}\else
  \providecommand{\doi}{doi:\discretionary{}{}{}\begingroup
  \urlstyle{rm}\Url}\fi
\providecommand{\bibAnnoteFile}[1]{%
  \IfFileExists{#1}{\begin{quotation}\noindent\textsc{Key:} #1\\
  \textsc{Annotation:}\ \input{#1}\end{quotation}}{}}
\providecommand{\bibAnnote}[2]{%
  \begin{quotation}\noindent\textsc{Key:} #1\\
  \textsc{Annotation:}\ #2\end{quotation}}
\providecommand{\eprint}[2][]{\url{#2}}

\bibitem{wright:Genetics:1931}
Wright S (1931) Evolution in {M}endelian populations.
\newblock Genetics 16: 97--159.
\bibAnnoteFile{wright:Genetics:1931}

\bibitem{kimura:Genetics:1964b}
Kimura M, Weiss G (1964) The stepping stone model of population structure and
  the decrease of genetic correlation with distance.
\newblock Genetics 49: 561-575.
\bibAnnoteFile{kimura:Genetics:1964b}

\bibitem{levins:BSEA:1969}
Levins R (1969) Some demographic and genetic consequences of environmental
  heterogeneity for biological control.
\newblock Bulletin of the Entomological Society of America 15: 237-240.
\bibAnnoteFile{levins:BSEA:1969}

\bibitem{nowak:nature:1992b}
Nowak MA, May RM (1992) Evolutionary games and spatial chaos.
\newblock Nature 359: 826-829.
\bibAnnoteFile{nowak:nature:1992b}

\bibitem{lieberman:nature:2005}
Lieberman E, Hauert C, Nowak MA (2005) Evolutionary dynamics on graphs.
\newblock Nature 433: 312-316.
\bibAnnoteFile{lieberman:nature:2005}

\bibitem{frucht:CJM:39}
Frucht R (1949) Graphs of degree three with a given abstract group.
\newblock Canadian Journal of Mathematics 1: 365-378.
\bibAnnoteFile{frucht:CJM:39}

\bibitem{ohtsuki:PRSB:2006}
Ohtsuki H, Nowak MA (2006) Evolutionary games on cycles.
\newblock Proceedings of the Royal Society B 273: 2249--2256.
\bibAnnoteFile{ohtsuki:PRSB:2006}

\bibitem{taylor:Nature:2007}
Taylor PD, Day T, Wild G (2007) Evolution of cooperation in a finite
  homogeneous graph.
\newblock Nature 447: 469-472.
\bibAnnoteFile{taylor:Nature:2007}

\bibitem{grafen:JTB:2008}
Grafen A, Archetti M (2008) Natural selection of altruism in inelastic viscous
  homogeneous populations.
\newblock Journal of Theoretical Biology 252: 694--710.
\bibAnnoteFile{grafen:JTB:2008}

\bibitem{ohtsuki:Nature:2006}
Ohtsuki H, Hauert C, Lieberman E, Nowak MA (2006) A simple rule for the
  evolution of cooperation on graphs.
\newblock Nature 441: 502-505.
\bibAnnoteFile{ohtsuki:Nature:2006}

\bibitem{dawes:ARP:1980}
Dawes RM (1980) Social dilemmas.
\newblock Annual Review of Psychology 31: 169-193.
\bibAnnoteFile{dawes:ARP:1980}

\bibitem{hauert:JTB:2006a}
Hauert C, Michor F, Nowak MA, Doebeli M (2006) Synergy and discounting of
  cooperation in social dilemmas.
\newblock Journal of Theoretical Biology 239: 195-202.
\bibAnnoteFile{hauert:JTB:2006a}

\bibitem{taylor:MB:1978}
Taylor PD, Jonker L (1978) Evolutionary stable strategies and game dynamics.
\newblock Mathematical Biosciences 40: 145-156.
\bibAnnoteFile{taylor:MB:1978}

\bibitem{hofbauer:book:1998}
Hofbauer J, Sigmund K (1998) Evolutionary Games and Population Dynamics.
\newblock Cambridge University Press, Cambridge.
\bibAnnoteFile{hofbauer:book:1998}

\bibitem{nowak:Nature:2004}
Nowak MA, Sasaki A, Taylor C, Fudenberg D (2004) Emergence of cooperation and
  evolutionary stability in finite populations.
\newblock Nature 428: 646-650.
\bibAnnoteFile{nowak:Nature:2004}

\bibitem{nowak:AAM:1990}
Nowak MA, Sigmund K (1990) The evolution of stochastic strategies in the
  prisoner's dilemma.
\newblock Acta Applicandae Mathematicae 20: 247-265.
\bibAnnoteFile{nowak:AAM:1990}

\bibitem{van-baalen:JTB:1998}
Van~Baalen M, Rand DA (1998) The unit of selection in viscous populations and
  the evolution of altruism.
\newblock Journal of Theoretical Biology 193: 631-648.
\bibAnnoteFile{van-baalen:JTB:1998}

\bibitem{hauert:PRSB:2001}
Hauert C (2001) Fundamental clusters in spatial $2\times2$ games.
\newblock Proceedings of the Royal Society B 268: 761-9.
\bibAnnoteFile{hauert:PRSB:2001}

\bibitem{fletcher:PRSB:2009}
Fletcher JA, Doebeli M (2009) A simple and general explanation for the
  evolution of altruism.
\newblock Proceedings of the Royal Society B 276: 13--19.
\bibAnnoteFile{fletcher:PRSB:2009}

\bibitem{zukewich:PlosOne:2013}
Zukewich J, Kurella V, Doebeli M, Hauert C (2013) Consolidating birth-death and
  death-birth processes in structured populations.
\newblock PLoS One 8: e54639.
\bibAnnoteFile{zukewich:PlosOne:2013}

\bibitem{maciejewski:JTB:2013}
Maciejewski W (2014) Reproductive value in graph-structured populations.
\newblock Journal of Theoretical Biology 340: 285-293.
\bibAnnoteFile{maciejewski:JTB:2013}

\bibitem{broom:JSTP:2011}
Broom M, Rychtar J, Stadler B (2011) Evolutionary dynamics on graphs - the
  effect of graph structure and initial placement on mutant spread.
\newblock Journal of Statistical Theory and Practice 5: 369-381.
\bibAnnoteFile{broom:JSTP:2011}

\bibitem{li:PLOS:2013}
Li C, Zhang B, Cressman R, Tao Y (2013) Evolution of cooperation in a
  heterogeneous graph: Fixation probabilities under weak selection.
\newblock PLoS One 8.
\bibAnnoteFile{li:PLOS:2013}

\bibitem{santos:PRL:2005}
Santos FC, Pacheco JM (2005) Scale-free networks provide a unifying framework
  for the emergence of cooperation.
\newblock Physical Review Letters 95: 098104.
\bibAnnoteFile{santos:PRL:2005}

\bibitem{santos:PRSB:2006}
Santos FC, Rodrigues JF, Pacheco JM (2006) Graph topology plays a determinant
  role in the evolution of cooperation.
\newblock Proceedings of the Royal Society B 273: 51-55.
\bibAnnoteFile{santos:PRSB:2006}

\bibitem{santos:PlosCB:2006}
Santos FC, Pacheco JM, Lenaerts T (2006) Cooperation prevails when individuals
  adjust their social ties.
\newblock PLoS Computational Biology 2: 1284-1291.
\bibAnnoteFile{santos:PlosCB:2006}

\bibitem{santos:Nature:2008}
Santos FC, Santos MD, Pacheco JM (2008) Social diversity promotes the emergence
  of cooperation in public goods games.
\newblock Nature 454: 213--216.
\bibAnnoteFile{santos:Nature:2008}

\bibitem{tomassini:IJMP:2007}
Tomassini M, Pestelacci E, Luthi L (2007) Social dilemmas and cooperation in
  complex networks.
\newblock International Journal of Modern Physics C 18: 1173-1185.
\bibAnnoteFile{tomassini:IJMP:2007}

\bibitem{szolnoki:PA:2008}
Szolnoki A, Perc M, Danku Z (2008) Towards effective payoffs in the prisoner's
  dilemma game on scale-free networks.
\newblock Physica A 387: 2075--2082.
\bibAnnoteFile{szolnoki:PA:2008}

\bibitem{antonioni:ACS:2012}
Antonioni A, Tomassini M (2012) Cooperation on social networks and its
  robustness.
\newblock Advances in Complex Systems 15.
\bibAnnoteFile{antonioni:ACS:2012}

\bibitem{huberman:PNAS:1993}
Huberman BA, Glance NS (1993) Evolutionary games and computer simulations.
\newblock Proceedings of the National Academy of Sciences USA 90: 7716-7718.
\bibAnnoteFile{huberman:PNAS:1993}

\bibitem{masuda:PRSB:2007a}
Masuda N (2007) Participation costs dismiss the advantage of heterogeneous
  networks in evolution of cooperation.
\newblock Proceedings of the Royal Society B 274: 1815--1821.
\bibAnnoteFile{masuda:PRSB:2007a}

\bibitem{perc:PRE:2008a}
Perc M, Szolnoki A (2008) Social diversity and promotion of cooperation in the
  spatial prisoner's dilemma game.
\newblock Physical Review E 77: 0011904.
\bibAnnoteFile{perc:PRE:2008a}

\bibitem{pacheco:PLoSCB:2009}
Pacheco J, Pinheiro FL, Santos FC (2009) Population structure induces a
  symmetry breaking favouring the emergence of cooperation.
\newblock PLoS Computational Biology 5: e1000596.
\bibAnnoteFile{pacheco:PLoSCB:2009}

\bibitem{grilo:JTB:2011}
Grilo C, Correia L (2011) Effects of asynchronism on evolutionary games.
\newblock Journal of Theoretical Biology 269: 109-122.
\bibAnnoteFile{grilo:JTB:2011}

\bibitem{hauert:Nature:2004}
Hauert C, Doebeli M (2004) Spatial structure often inhibits the evolution of
  cooperation in the snowdrift game.
\newblock Nature 428: 643-646.
\bibAnnoteFile{hauert:Nature:2004}

\bibitem{hauert:IJBC:2002}
Hauert C (2002) Effects of space in $2\times2$ games.
\newblock International Journal of Bifurcation and Chaos 12: 1531-1548.
\bibAnnoteFile{hauert:IJBC:2002}

\bibitem{szabo:PR:2007}
Szab{\'o} G, F{\'a}th G (2007) Evolutionary games on graphs.
\newblock Physics Reports 446: 97-216.
\bibAnnoteFile{szabo:PR:2007}

\bibitem{barabasi:Science:1999}
Barab{\'{a}}si A, Albert R (1999) Emergence of scaling in random networks.
\newblock Science 286: 509-512.
\bibAnnoteFile{barabasi:Science:1999}

\bibitem{santos:JTB:2012}
Santos FC, Pinheiro FL, Lenaerts T, Pacheco JM (2012) The role of diversity in
  the evolution of cooperation.
\newblock Journal of Theoretical Biology 299: 88-96.
\bibAnnoteFile{santos:JTB:2012}

\bibitem{traulsen:PRL:2005}
Traulsen A, Claussen JC, Hauert C (2005) Coevolutionary dynamics: From finite
  to infinite populations.
\newblock Physical Review Letters 95: 238701.
\bibAnnoteFile{traulsen:PRL:2005}

\bibitem{traulsen:PRE:2012}
Traulsen A, Claussen JC, Hauert C (2012) Stochastic differential equations for
  evolutionary dynamics with demographic noise and mutations.
\newblock Physical Review E 85: 041901.
\bibAnnoteFile{traulsen:PRE:2012}

\bibitem{szabo:PRE:1998}
Szab{\'o} G, T{\H o}ke C (1998) Evolutionary {P}risoner's {D}ilemma game on a
  square lattice.
\newblock Physical Review E 58: 69-73.
\bibAnnoteFile{szabo:PRE:1998}

\bibitem{taylor:JTB:2007}
Taylor PD, Day T, Wild G (2007) From inclusive fitness to fixation probability
  in homogeneous structured populations.
\newblock Journal of Theoretical Biology 249: 101-110.
\bibAnnoteFile{taylor:JTB:2007}

\bibitem{hadjichrysanthou:DGA:2011}
Hadjichrysanthou C, Broom M, Rycht\'{a}\v{r} J (2011) Evolutionary games on
  star graphs under various updating rules.
\newblock Dynamic Games and Applications 1: 386-407.
\bibAnnoteFile{hadjichrysanthou:DGA:2011}

\bibitem{woelfing:JTB:2009}
Woelfing B, Traulsen A (2009) Stochastic sampling of interaction partners
  versus deterministic payoff assignment.
\newblock Journal of Theoretical Biology 257: 689--695.
\bibAnnoteFile{woelfing:JTB:2009}

\bibitem{szabo:PRE:2005}
Szab{\'o} G, Vukov J, Szolnoki A (2005) Phase diagrams for an evolutionary
  prisoner's dilemma game on two-dimensional lattices.
\newblock Physical Review E 72: 047107.
\bibAnnoteFile{szabo:PRE:2005}

\bibitem{pinheiro:PLOS:2012}
Pinheiro F, Pacheco JM, Santos F (2012) From local to global dilemmas in social
  networks.
\newblock PLoS One 7(2): e32114.
\bibAnnoteFile{pinheiro:PLOS:2012}

\bibitem{pinheiro:NJP:2012}
Pinheiro FL, Santos FC, Pacheco JM (2012) How selection pressure changes the
  nature of social dilemmas in structured populations.
\newblock New Journal of Physics 14: 073035.
\bibAnnoteFile{pinheiro:NJP:2012}

\bibitem{erdos:PMHAS:1960}
Erd{\H{o}}s P, R{\'{e}}nyi A (1960) On the evolution of random graphs.
\newblock Publ Math Inst Hung Acad Sci 5: 17-61.
\bibAnnoteFile{erdos:PMHAS:1960}

\bibitem{newman:PRE:1999}
Newman MEJ, Watts DJ (1999) Scaling and percolation in the small-world network
  model.
\newblock Physical Review E 60: 7332.
\bibAnnoteFile{newman:PRE:1999}

\bibitem{klemm:PRE:2002a}
Klemm K, Eguiluz VM (2002) Highly clustered scale-free networks.
\newblock Physical Review E 65: 036123.
\bibAnnoteFile{klemm:PRE:2002a}

\end{thebibliography}

\section*{Figure Legends}

\end{document}